\documentclass[12pt,fleqn,draft]{article}
\usepackage{epsf,mcite}
\makeatletter
\@addtoreset{equation}{section}
\renewcommand\theequation{\thesection.\arabic{equation}}
\renewcommand\appendix{\par
  \setcounter{section}{0}
  \renewcommand\thesection{Appendix \Alph{section}}
  \renewcommand\theequation{\Alph{section}\arabic{equation}}}
\renewenvironment{titlepage}{
  \pagestyle{empty}}
  {\newpage\pagestyle{plain}\setcounter{page}{1}}
\long\def\@makecaption#1{%
  \vskip\abovecaptionskip
  \sbox\@tempboxa{#1}%
  \ifdim \wd\@tempboxa >\hsize
    #1\par
  \else
    \hbox to\hsize{\hfil\box\@tempboxa\hfil}%
  \fi
  \vskip\belowcaptionskip}

\def\@bibitem#1{%
  \ifmc@bstsupport
    \mc@iftail{#1}%
      {;\\ \space\ignorespaces}%
      {\ifmc@first\else.\fi\orig@bibitem{#1}}
    \mc@firstfalse
  \else
    \mc@iftail{#1}%
      {\ignorespaces}%
      {\orig@bibitem{#1}}%
  \fi}%
\newif\ifmc@bstsupport
\mc@bstsupportfalse
\newcommand{\pub}[2]{
  \hfill INLO--PUB--#1/#2
  \vspace*{1.5cm}}
\newcommand{\institute}{
  Instituut--Lorentz\\
  University of Leiden\\
  P.O. Box 9506\\
  2300 RA Leiden\\
  The Netherlands}
\renewcommand{\title}[3]{
  \begin{center}
    \large{\bf #1}\\[1.8cm]
    #2\\[1.8cm]
    \institute\\[1.5cm]
    #3\vspace*{2cm}
  \end{center}}
\renewcommand{\abstract}[1]{\noindent {\bf Abstract}\\[8mm] \noindent #1}
\newenvironment{mytable}[3]
  {\begin{table}\caption[]{\label{#2}#3}\vspace*{2ex}
   \begin{center}\begin{tabular}{#1}}
  {\end{tabular}\end{center}\end{table}}
\newcommand{\citegroup}[3]{\cite{#1}\vphantom{\cite{#2}}-\cite{#3}}
\newcommand{\etal}{{\em et al.}}
\newcommand{\eref}[1]{(\ref{#1})}
\newcommand{\MS}{$\overline{\mbox{MS}}$}
\newcommand{\BigLeftHook}{\left[\vphantom{\frac{A}{A}}\right.}
\newcommand{\BigRightHook}{\left.\vphantom{\frac{A}{A}}\right]}
\newcommand{\BigLeftBrace}{\left\{\vphantom{\frac{A}{A}}\right.}
\newcommand{\BigRightBrace}{\left.\vphantom{\frac{A}{A}}\right\}}
\newcommand{\BigLeftParen}{\left(\vphantom{\frac{A}{A}}\right.}
\newcommand{\BigRightParen}{\left.\vphantom{\frac{A}{A}}\right)}
\newcommand{\Li}{{\rm Li}}
\newcommand{\calO}{{\cal O}}
\newcommand{\alphas}{$\alpha_s$}
\newcommand{\alphastwo}{$\alpha_s^2$}
\newcommand{\ten}[1]{\cdot 10^{#1}}
\input{./library.ps}
\newcommand{\diagram}[1]{\special{" ProcDict begin #1 end}}
\makeatother
\begin{document}
\begin{titlepage}
\pub{1}{95}
\title{Heavy flavor contributions to the Drell-Yan cross section}
      {P.J. Rijken and W.L. van Neerven}
      {January 1995}
\abstract{
We investigate the effect of heavy flavor contributions to
vector boson ($V$ = $\gamma$, $Z$, $W$) production which is described by
the Drell-Yan
mechanism. All reactions with bottom and top quarks ($Q_i$ = $b$, $t$)
in the final state, like
$q_1 + \bar{q}_2\rightarrow V + Q_1 + \bar{Q}_2$ and
$g + g\rightarrow V + Q_1 + \bar{Q}_2$, are considered.
This study also includes the
virtual contributions containing heavy flavor loops which were not taken
into account earlier in the literature. Our analysis reveals that the above
corrections to the Drell-Yan cross section are very small. Only
at energies characteristic for the LHC they are of the
same order of magnitude as the order \alphastwo~QCD contributions
due to light quark and gluon subprocesses calculated earlier in the
literature.}
\end{titlepage}
\newpage
\section{Introduction}
The production cross sections for the electroweak vector bosons
$W$ and $Z$ in hadron-hadron collisions as described by the Drell-Yan
(DY)~\cite{Dre70} mechanism
provides us with a beautiful test of perturbative QCD.
One of the reasons is that the total cross section can be calculated
in next-to-next-to leading order in perturbative QCD \cite{Ham91,Nee92}
a result
which is very hard to achieve for other processes in hadron-hadron
collisions.\\
Another advantage of this process is that on the Born level it is purely
electroweak in origin for which the theory is in an excellent shape. Hence
each deviation of the cross section from the Born approximation can be
attributed to QCD effects. Therefore the Drell-Yan (DY) process belongs
to the same class as deep inelastic lepton hadron scattering and
electron-positron collisions where QCD corrections can be measured
with much higher accuracy than in other reactions.\\
Besides the total cross section for vector boson production and the
cross section $d\sigma/dm$ where $m$ denotes the lepton pair invariant
mass order \alphastwo~corrections due to soft plus virtual gluon
contributions have been calculated in \cite{Mat88a,*Mat89,*Rij94}
to the differential
distributions $d^2\sigma/dmdx_F$ and $d^2\sigma/dmdy$. Here the quantities
$x_F$ and $y$ denote the fraction of the longitudinal momentum of the
lepton pair with respect to the center of mass (CM) momentum and the
rapidity respectively.\\
Furthermore one has computed the full order \alphas~correction to the
single vector boson inclusive cross sections $d^3\sigma/dmdx_Fdp_T$
or $d^3\sigma/dmdydp_T$~\cite{Arn89,*Gon89} where $p_T$ denotes the transverse
momentum of the vector boson. All the above calculations have been
performed under the assumption that the quarks, appearing in the partonic
subprocesses contributing to the DY process, are massless. This is a
reasonable assumption for the light quarks $u$, $d$, and $s$ and even for
$c$ since the masses of the $W$ and $Z$ are large compared with the masses
of the above quarks. However this assumption is doubtful for the bottom
quark and it is certainly wrong for the top quark since recent experiments
\cite{Abe94} indicate that $m_t > M_W$, $M_Z$. Therefore we cannot neglect
the masses of the bottom and top in the final state of the partonic
subprocesses in particular if the collider CM energy is small like in
the case of the $Sp\bar{p}S$ ($\sqrt{s} = 0.63\,{\rm TeV}$) or the
TEVATRON ($\sqrt{s} = 1.8\,{\rm TeV}$). Maybe for some partonic subprocesses
one can apply the zero mass
approximation for the bottom quark at LHC ($\sqrt{s}=16\,{\rm TeV}$) as we
will show later on. Contrary to the final
state  we will omit the bottom and top quarks in the initial state because
we assume that the bottom and top densities in the hadron are
negligibly small.\\
The calculation of the contribution of heavy flavors to the DY cross
section has been performed for $Z$-production in \cite{Dic86}. It
contained all one-loop and two-loop corrections which are characteristic
of $Z$-production but do not contribute to $W$-production or processes
with a photon in the  intermediate state. They are characterized by the
triangular heavy flavor-loop insertions containing the Adler-Bell-Jackiw
anomaly which has to cancel while adding top and bottom loops. This work
was extended in \cite{Gon92} by including the interference terms
originating from the final and initial state radiation of the vector
boson in the process $q+\bar{q}\rightarrow Z+Q+\bar{Q}$ (with $Q$ =
$b$, $t$). All other production mechanisms, which also show up for
$W$-production and processes with a photon in the intermediate
state, have not been considered yet. The contributions to $Z$-production
considered in \cite{Dic86,Gon92} all show up on the order \alphastwo~
level and amount to about 0.1 percent of the Born approximation which
means that they are experimentally unobservable.\\
In this paper we want to complete the above analysis by including all
remaining Feynman graphs which also contribute to lepton pair production
with a $W$ or a photon in the intermediate state. Apart of some additional
two-loop graphs they contain the contribution of the partonic subprocesses
$q_1 + \bar{q}_2\rightarrow V + Q_1+ \bar{Q}_2$ and $g + g\rightarrow V + Q_1
+\bar{Q}_2$ with
$V$ = $\gamma$, $Z$, $W$ and $Q_i$ = $b$, $t$. Like the corrections discussed
in~\cite{Dic86,Gon92} they all contribute to the DY cross section on the
order \alphastwo~level.\\
This paper will be organized as follows. In section 2 we present the
partonic cross sections corresponding to the subprocesses which
contribute to heavy flavors plus vector boson production. Furthermore
we show that in the case the vector boson mass becomes much larger
than the heavy flavor mass one can find explicit analytic results. In
section 3 we compute the heavy flavor contributions to vector boson
production at current and future hadron-hadron colliders. In particular
a comparison will be made between the order \alphastwo~corrections due
to light partons (quarks and gluons), calculated previously in
\cite{Ham91,Nee92}, and the contributions due to bottom and top quarks
presented in this paper. Finally in Appendix A we give an explicit
formula for the two-loop vertex correction which was not computed in
\cite{Dic86,Gon92}.
\newpage
\section{The order \alphastwo~corrections to the DY cross section
         due to heavy flavor production}
In this section we present the partonic cross sections of heavy
flavor production contributing to the Drell-Yan (DY) process
which is given by
\begin{eqnarray}
  \label{eq:DY_process}
  H_1 + H_2 &\rightarrow& V + ``X" \nonumber\\
  && \lfloor\hspace*{-2pt}\raisebox{-5.7pt}{$\rightarrow \ell_1 + \ell_2$,}
\end{eqnarray}
where $H_1$, $H_2$ denote the incoming hadrons and $V$ stands for one
of the vector bosons of the standard electroweak model ($\gamma$, $Z$
or $W$) which subsequently decays in the lepton pair $\ell_1$, $\ell_2$.\\
The heavy flavors in the final state are given by $Q_1$ and $Q_2$
respectively and the symbol $``X"$ denotes any inclusive final hadronic
state. In lowest order of the electroweak and strong coupling constants
the above reaction receives contributions of the following parton
subprocesses
\begin{equation}
  \label{eq:parton_process}
  i(k_1) + j(k_2) \rightarrow V(q) + Q_1(p_1) + \bar{Q}_2(p_2),
\end{equation}
with $i$, $j$ = $q$, $\bar{q}$, $g$. Here $q$ stands for the light quarks
given by $u$, $d$, $s$, and $c$ whereas the heavy quarks are represented
by $t$ and $b$. Notice that in this paper we study heavy flavor production
at large hadron collider energies so that the charm can be treated
as a light quark which mass can be neglected. In addition to reaction
\eref{eq:parton_process} we include the virtual corrections due to heavy
flavor loops which contribute to the subprocesses
\begin{eqnarray}
  \label{eq:sub_V}
  i(k_1) + j(k_2) &\rightarrow& V(q), \\[2ex]
  \label{eq:sub_Vk}
  i(k_1) + j(k_2) &\rightarrow& V(q) + l(k_3),
\end{eqnarray}
with $l = q$, $\bar{q}$, $g$. The most of these virtual corrections
reveal the presence of the triangle fermion loops giving rise to the
well known Adler-Bell-Jackiw anomaly which has to cancel between the top
and bottom contributions. They have been treated in \cite{Dic86,Gon92}
in the
case of $V = Z$. Here below we will add the contributions due to gluon
self energies containing the heavy quark loop which also appear when $V$
is represented by $\gamma$ and $W$.\\
In the subsequent part of this paper we are interested in the following
parton cross section
\begin{equation}
  \label{eq:parton_cross}
  \frac{d\hat{\sigma}_{ij}^V}{dQ^2} = \hat{\tau}\,\sigma_V(Q^2, M_V^2)\,
  \hat{W}_{ij}^V(\hat{\tau},Q^2,m_1^2,m_2^2),\hspace*{2cm}
  \hat{\tau} = \frac{Q^2}{\hat{s}},
\end{equation}
receiving  contributions from the reactions
\eref{eq:parton_process}-\eref{eq:sub_Vk}
($i$, $j$ = $q$, $\bar{q}$, $g$). The quantity
$\sigma_V$ represents the pointlike DY cross section and the kinematical
variables $\sqrt{\hat{s}}$ ($\hat{s} = (k_1 + k_2)^2$) and $\sqrt{Q^2}$
stand for the C.M. energy and the lepton pair invariant mass respectively.
In addition to the above variables the DY structure function,
denoted by $\hat{W}_{ij}^V$, also depends on the heavy flavor masses $m_1$ and
$m_2$. The pointlike cross section refers to the reaction
\begin{equation}
  \label{eq:pointlike_cross_section}
  q_1 + \bar{q}_2 \rightarrow V \rightarrow \ell_1 + \bar{\ell}_2,
\end{equation}
and is given by (see \cite{Ham91,Nee92})
\begin{eqnarray}
  \label{eq:point_gamma}
  \sigma_\gamma(Q^2) &=& \frac{1}{N}\,\frac{4\pi\alpha^2}{3Q^4},\\[2ex]
  \label{eq:point_Z}
  \sigma_Z(Q^2,M_Z^2) &=& \frac{1}{N}\,\frac{\pi\alpha}
    {4M_Z\sin^2\theta_W\cos^2\theta_W}\,\frac{\Gamma_{Z\rightarrow
    \ell\bar{\ell}}}{(Q^2-M_Z^2)^2 + M_Z^2\Gamma_Z^2},\\[2ex]
  \label{eq:point_W}
  \sigma_W(Q^2,M_W^2) &=& \frac{1}{N}\,\frac{\pi\alpha}
    {M_W\sin^2\theta_W}\,\frac{\Gamma_{W\rightarrow
    \ell\bar{\nu}}}{(Q^2-M_W^2)^2 + M_W^2\Gamma_W^2}.
\end{eqnarray}
For completeness we also give the formula for the $\gamma-Z$ interference
\begin{equation}
  \label{eq:interference}
  \sigma_{\gamma Z}(Q^2,M_Z^2) = \frac{1}{N}\,\frac{\pi\alpha^2}{6}\,
  \frac{1-4\sin^2\theta_W}{\sin^2\theta_W\cos^2\theta_W}\,\frac{1}{Q^2}
  \,\frac{Q^2-M_Z^2}{(Q^2-M_Z^2)^2+M_Z^2\Gamma_Z^2}.
\end{equation}
In the above expression $\theta_W$ denotes the electroweak angle.
Furthermore $\Gamma_Z$ and $\Gamma_W$ represent the total width of the
$Z$ and $W$ boson respectively (sum over all decay channels) and
$N=3$ ($SU(N)$, color). The partial widths due to the leptonic decay
of the $Z$ and $W$ are given by
\begin{eqnarray}
  \label{eq:decay_Z}
  \Gamma_{Z\rightarrow\ell\bar{\ell}} &=&
    \frac{\alpha M_Z(1 + (1-4\sin^2\theta_W)^2)}
         {48\sin^2\theta_W\cos^2\theta_W}, \\[2ex]
  \label{eq:decay_W}
  \Gamma_{W\rightarrow\ell\bar{\nu}_\ell} &=&
    \frac{\alpha M_W}{12\sin^2\theta_W}.
\end{eqnarray}
In the case of $W$- and $Z$-production the total cross section can be
obtained using the narrow width approximation while integrating
\eref{eq:parton_cross} over $Q^2$, i.e.
\begin{equation}
  \label{eq:narrow_width_V}
  \frac{1}{(Q^2-M_V^2)^2 + M_V^2\Gamma_V^2}\rightarrow
    \frac{\pi}{M_V\Gamma_V}\,\delta(Q^2-M_V^2).
\end{equation}
The total rates $\hat{\sigma}_{ij}^V$
(sum over all leptonic and hadronic decay channels) are now given by
\begin{eqnarray}
  \label{eq:total_parton_Z}
  \hat{\sigma}_{ij}^Z &=& \frac{1}{N}\,\frac{\pi^2\alpha}
  {4\sin^2\theta_W\cos^2\theta_W}\,\frac{1}{\hat{s}}\,
  \hat{W}_{ij}^Z(M_Z^2/\hat{s},M_Z^2,m^2,m^2),\\[2ex]
  \label{eq:total_parton_W}
  \hat{\sigma}_{ij}^W &=& \frac{1}{N}\,\frac{\pi^2\alpha}
  {\sin^2\theta_W}\,\frac{1}{\hat{s}}\,
  \hat{W}_{ij}^W(M_W^2/\hat{s},M_W^2,m_1^2,m_2^2).
\end{eqnarray}
When we consider the reaction where the vector boson decays into a
specific lepton pair $\ell_1$, $\bar{\ell}_2$~~$\hat{\sigma}_{ij}^V$
in expressions
\eref{eq:total_parton_Z}, \eref{eq:total_parton_W} has to be replaced by
\begin{equation}
  \label{eq:cross_branch}
  \hat{\sigma}_{ij}^{V\rightarrow\ell_1\bar{\ell}_2} =
    \hat{\sigma}_{ij}^V\,B(V\rightarrow \ell_1\bar{\ell}_2),
\end{equation}
where $B(V\rightarrow\ell_1\bar{\ell}_2)$ stands for the branching
ratio
\begin{equation}
  \label{eq:branching_ratio}
  B(V\rightarrow\ell_1\bar{\ell}_2) = \frac{\Gamma_{V\rightarrow
  \ell_1\bar{\ell}_2}}{\Gamma_V}.
\end{equation}
Notice that all particles into which the vector bosons decay are taken
to be massless which is a good approximation since $M_Z >> m_b$ and the
top is too heavy to appear in the decay products.\\
Further since the electroweak radiative corrections to $\sin^2\theta_W$ are
non negligible it is better to replace $\sin^2\theta_W$ appearing in the
denominators of the above expressions by
\begin{equation}
  \label{eq:sin}
  \sin^2\theta_W = \frac{\pi\alpha}{G_F\sqrt{2}M_W^2},
\end{equation}
where $G_F = 1.1667\ten{-5}\,{\rm GeV}^{-2}$ (Fermi constant) whereas
in the numerators we have put $\sin^2\theta_W = 0.2258$.\\
The above definitions imply that the vector- and axial-vector couplings
describing the strength of the coupling of the vector bosons to
the quarks are hidden
in the definition for $\hat{W}_{ij}^V$ in \eref{eq:parton_cross}. The same
also holds for the elements of the Kobayashi-Maskawa matrix,
denoted by $V_{q_1\bar{q}_2}$, which
we approximate by retaining the Cabibbo angles only and putting the
remaining angles and phases equal to zero.\\
In the case of the Born process as given by reaction
\eref{eq:pointlike_cross_section}
the DY-structure function becomes
\begin{equation}
  \label{eq:W_qbarq_V}
  \hat{W}_{q\bar{q}}^V = \left|V_{q_1\bar{q}_2}\right|^2
                         \left( \left(v_q^V\right)^2
    + \left(a_q^V\right)^2 \right)\,\delta(1-\hat{\tau}),
\end{equation}
with $V_{q_1\bar{q}_2} = 1$ in the case $V$ = $\gamma$, $Z$.
The vector- and axial-vector couplings are equal to
\begin{eqnarray}
  \label{eq:couplings}
  v_u^\gamma = \frac{2}{3}, && a_u^\gamma = 0, \nonumber\\[2ex]
  v_d^\gamma = -\frac{1}{3}, && a_d^\gamma = 0, \nonumber\\[2ex]
  v_u^Z = 1 - \frac{8}{3}\sin^2\theta_W, && a_u^Z = -1, \nonumber\\[2ex]
  v_d^Z = -1 + \frac{4}{3}\sin^2\theta_W, && a_d^Z = 1, \nonumber\\[2ex]
  v_u^W = v_d^W = \frac{1}{\sqrt{2}}, && a_u^W = a_d^W = -\frac{1}{\sqrt2}.
\end{eqnarray}
We will now list all order \alphastwo~parton cross sections due to heavy
flavor production contributing to reaction \eref{eq:DY_process}.\\
We start with the two-loop corrections to the Born process
\begin{equation}
  \label{eq:sub_qbarq_V}
  q_1 + \bar{q}_2 \rightarrow V,
\end{equation}
which are presented by the graphs in Figs.~\ref{fig:W_qbarq_Z_2} and
\ref{fig:W_qbarq_V_2}. The first one (Fig.~\ref{fig:W_qbarq_Z_2}) contains
a heavy quark in the triangular loop and it only contributes to
$Z$-production. The $W$ does not contribute because of charge conservation
and the same holds for the photon due to charge conjugation
(Furry's theorem). The DY structure function is equal to
\citegroup{Dic86}{Gon92}{Kni90}
\begin{eqnarray}
  \label{eq:W_qbarq_Z_2}
  \lefteqn{W_{q\bar{q}}^Z = \delta(1-\hat{\tau})\,a_q^Za_Q^Z\,C_F T_f\,
  \frac{1}{2}
  \left(\frac{\alpha_s}{\pi}\right)^2\BigLeftHook}\nonumber\\[2ex]
  && \theta(Q^2-4m^2)G_1(m^2/Q^2) + \theta(4m^2-Q^2)G_2(m^2/Q^2)
  \BigRightHook,
\end{eqnarray}
where $C_F$ and $T_f$ are the $SU(N)$ color factors: $C_F =
\frac{N^2-1}{2N}$, $T_f = \frac{1}{2}$ and $q = u$, $d$, $s$, $c$~~~,
$Q = b$, $t$. The functions $G_1$ and $G_2$ are given in Eqs. (2.8) and
(2.9) of \cite{Gon92} respectively.
Notice that in expression \eref{eq:W_qbarq_Z_2} one has to sum over $b$ and
$t$ in order to cancel the Adler-Bell-Jackiw axial anomaly.\\
The second two-loop correction to \eref{eq:sub_qbarq_V} is given by
the vertex correction in Fig.~\ref{fig:W_qbarq_V_2}. It contains the heavy
quark loop contribution to the gluon self energy insertion in the
vertex graph. The DY structure function becomes
\begin{equation}
  \label{eq:W_qbarq_V_2}
  W_{q\bar{q}}^V = \delta(1-\hat{\tau})\,
           \left|V_{q_1\bar{q}_2}\right|^2\,
           \left(\left(v_q^V\right)^2+\left(a_q^V\right)^2\right)\,
            C_F T_f\,\frac {1}{8}\left(\frac{\alpha_s}{\pi}\right)^2\,
            F(Q^2,m^2),
\end{equation}
where $F(Q^2,m^2)$ can be found in \eref{eq:F_qbarq_V_2}. The above expression
contributes for all vector bosons and  $V_{q_1\bar{q}_2} = 1$ when
$V$ = $\gamma$, $Z$.\\
Next we have the one-loop corrections to the two-to-two body processes
(see Figs.~\ref{fig:W_qbarq_Z_1} and \ref{fig:W_qg_Z})
\begin{eqnarray}
  \label{eq:sub_gV}
  q + \bar{q} &\rightarrow& g + V, \\[2ex]
  \label{eq:sub_qV}
  g + q(\bar{q}) &\rightarrow& q(\bar{q}) + V.
\end{eqnarray}
They all contain the triangle heavy quark loop insertion which
only contributes to $Z$-production for the same reasons as mentioned
above \eref{eq:W_qbarq_Z_2}. For process \eref{eq:sub_gV}
(Fig.~\ref{fig:W_qbarq_Z_1})
the DY structure function reads
\begin{eqnarray}
  \hat{W}_{q\bar{q}}^Z &=& a_q^Z a_Q^Z\,C_F T_f\,\frac{1}{2}
  \left(\frac{\alpha_s}{\pi}\right)^2\,\BigLeftHook
  \frac{1+\hat{\tau}}{1-\hat{\tau}}\,\left\{-2 + 2\hat{\tau}\left(
  J_1(4m^2/\hat{s}) - J_1(4m^2/Q^2)\right)\right\}\nonumber\\[2ex]
  \label{eq:W_qbarq_Z_1}
  && -\frac{4m^2}{\hat{s}}\left(J_2(4m^2/\hat{s}) - J_2(4m^2/Q^2)
  \right)\BigRightHook,
\end{eqnarray}
where $J_1$ and $J_2$ are given in Eqn. (2.12) of \cite{Gon92}.\\
For process \eref{eq:sub_qV} (Fig. \ref{fig:W_qg_Z}) we have the expression
\begin{eqnarray}
  \label{eq:W_qg_Z}
  \lefteqn{\hat{W}_{qg}^Z = a_q^Z a_Q^Z\,T_f^2\,\frac{1}{2}
  \left(\frac{\alpha_s}{\pi}\right)^2\,
  \BigLeftHook\theta(Q^2-4m^2)H_1(\hat{s},Q^2,m^2)}\nonumber\\[2ex]
  && \hspace*{1cm}+\theta(4m^2-Q^2)H_2(\hat{s},Q^2,m^2)\BigRightHook,
\end{eqnarray}
with $H_1$ and $H_2$ defined in (2.18) and (2.19) of \cite{Gon92} respectively.
Like for the DY structure function in \eref{eq:W_qbarq_Z_2} the above
expressions \eref{eq:W_qbarq_Z_1} and \eref{eq:W_qg_Z} have to be
summed over $b$ and $t$ in order to cancel the axial anomaly.\\
The next contributions are given by the two to three body reactions
(see Figs.~\ref{fig:W_qbarq_V_QbarQ} and \ref{fig:W_gg_V_QbarQ})
\begin{eqnarray}
  \label{eq:sub_qbarq_V_QbarQ}
  q(k_1) + \bar{q}(k_2) &\rightarrow&
    V(q) + Q_1(p_1) + \bar{Q}_2(p_2),\\[2ex]
  \label{eq:sub_gg_V_QbarQ}
  g(k_1) + g(k_2) &\rightarrow& V(q) + Q_1(p_1) + \bar{Q}_2(p_2).
\end{eqnarray}
Since at this moment experiments indicate that $m_t > m_b + M_W$ \cite{Abe94}
$W$-production has to be treated in a way which differs from the usual
procedure applied to $Z$- and $\gamma$-production. This is due to the
instability of the top quark appearing in the internal lines of the
Feynman graphs in Figs.~\ref{fig:W_qbarq_V_QbarQ} and
\ref{fig:W_gg_V_QbarQ} corresponding to the
reactions
\begin{eqnarray}
  \label{eq:qbarq_top_1}
  q + \bar{q} &\rightarrow& t + \bar{t} \\
  && \lfloor\hspace*{-2pt}\raisebox{-5.7pt}{$\rightarrow
    W^+ + b$,} \nonumber\\[2ex]
  \label{eq:qbarq_top_2}
  q + \bar{q} &\rightarrow& \bar{t} + t \\
  && \lfloor\hspace*{-2pt}\raisebox{-5.7pt}{$\rightarrow
    W^- + \bar{b}$,} \nonumber\\[2ex]
  \label{eq:gg_top_1}
  g + g &\rightarrow& t + \bar{t} \\
  && \lfloor\hspace*{-2pt}\raisebox{-5.7pt}{$\rightarrow
    W^+ + b$,} \nonumber\\[2ex]
  \label{eq:gg_top_2}
  g + g &\rightarrow& \bar{t} + t \\
  && \lfloor\hspace*{-2pt}\raisebox{-5.7pt}{$\rightarrow
    W^- + \bar{b}$.} \nonumber
\end{eqnarray}
In this case the internal top quark line cannot be described by an
ordinary Feynman propagator anymore.
Therefore it has to be treated as a resonance which implies that the
top quark propagator has to be replaced by a Breit- Wigner form. However
this procedure leads to a violation of gauge invariance which is
hard to remedy. In the case of the total cross section
\eref{eq:total_parton_W}
one can resort to an approximation which will be presented at the end
of this section.\\
Starting with $Z$- and $\gamma$-production which is described by the
Feynman diagrams in Figs.~\ref{fig:W_qbarq_V_QbarQ}
and \ref{fig:W_gg_V_QbarQ} the DY structure
function is given by
\begin{eqnarray}
  \label{eq:W_ij_V}
  \lefteqn{\hat{W}_{ij}^V = K_{ij}\frac{1}{8\pi}\,
  \left(\frac{\alpha_s}{4\pi}\right)^2\,
  \frac{1}{\hat{s}}\int d\hat{t}_1\,\int d\hat{u}_1\,
  \lambda^{1/2}\left(1,\frac{m_1^2}{\hat{s}_4},\frac{m_2^2}{\hat{s}_4}
  \right)\,}\nonumber\\[2ex]
  && \hspace*{3cm}\int_0^\pi d\phi\,\int_0^\pi d\theta\,\sin\theta\,
  \left| {\cal M}_{ij}^V\right|^2,
\end{eqnarray}
where ($i$, $j$) = ($q$, $\bar{q}$) or ($g$, $g$). The symbol $K_{ij}$
denotes the color factor which is given by $K_{q\bar{q}} = C_F T_f$ and
$K_{gg} = T_f^2$. In the definition of $\hat{W}_{ij}^V$ an average over
the initial spins and a sum over the final spins is understood. The
K\"allen function is defined by $\lambda(x,y,z) = x^2 + y^2 + z^2
- 2xy -2xz - 2yz$ and the kinematical invariants $\hat{s}_4$, $\hat{t}_1$,
and $\hat{u}_1$ are given by (see \eref{eq:sub_qbarq_V_QbarQ} and
\eref{eq:sub_gg_V_QbarQ})
\begin{equation}
  \label{eq:kin_inv}
  \hat{s}_4 = (p_1 + p_2)^2,\hspace*{1cm}
  \hat{t}_1 = (k_2 - q)^2,\hspace*{1cm}
  \hat{u}_1 = (k_1 - q)^2.
\end{equation}
In addition to the above invariants the matrix element squared
$\left|{\cal M}_{ij}^V\right|^2$ depends on other kinematical variables
which are analogous to the ones defined in Eqn. (4.2) of \cite{Bee89}. After
having performed the traces the computation of $\left|{\cal M}_{ij}^V
\right|^2$requires an intensive partial fractioning before the angular
integration can be carried out. The angular integrals can be found in
Appendix C of \cite{Bee89}. Notice that in the definition of $|{\cal M}|^2$
we have removed all the strong and electroweak coupling constants
described by $g_s$, $e$, and $g$ as defined in sections 10.6 and 14.5
of \cite{Bai93}. This also includes the typical factors which appear in the
vertices like $-ig_s\gamma_\mu T_a$, $-iev_q^\gamma\gamma_\mu$,
$-\frac{ig}{2}V_{q_1\bar{q}_2}\gamma_\mu(v_q^W + \gamma_5 a_q^W)$ and
$-\frac{ig}{4\cos\theta_W}\gamma_\mu(v_q^Z + \gamma_5 a_q^Z)$
(see Eqn. \eref{eq:couplings}). Here $V_{q_1\bar{q}_2}$ denotes the
Kobayashi-Maskawa matrixelement where only
the Cabibbo angle $\theta_C$ has been put to be unequal to zero.
All these factors are absorbed
in the definition of the pointlike cross sections $\sigma_V(Q^2,M_V^2)$
as presented in \eref{eq:point_gamma}-\eref{eq:point_W} except for the
Cabibbo angles and the couplings $v_q^V$ and $a_q^V$ \eref{eq:couplings}
which remain in $\hat{W}_{ij}^V$. In the case the vector boson couples
to massless quarks they can be factored out too like in
\eref{eq:W_qbarq_V} but for massive quarks like the heavy flavors
they remain in $\left|{\cal M}_{ij}^V\right|^2$.\\
In the case $Q^2 >> m^2$ ($m_1 = m_2 = m$) one can obtain analytic
expressions for the DY structure functions $\hat{W}_{q\bar{q}}^V$ and
$\hat{W}_{gg}^V$. For process \eref{eq:sub_qbarq_V_QbarQ} where the
vector boson is radiated off the incoming light quark lines
(see Figs.~\ref{fig:W_qbarq_V_QbarQ}a,b) one obtains
\footnote{The polylogarithms of the
type $\Li_n(x)$, $S_{n,p}(x)$ are defined in
\cite{Lew83,*Bar72,*Dev84}.}
\begin{eqnarray}
  \label{eq:W_qbarq_V_(1)}
  \lefteqn{W_{q\bar{q}}^{V,(1)}\left(\hat{\tau},\frac{Q^2}{m^2}
  \right) =
  \left|V_{q_1\bar{q}_2}\right|^2
  \left\{\left(v_q^V\right)^2+\left(
  a_q^V\right)^2\right\}\,C_F T_f\,\left(\frac{\alpha_s}{4\pi}\right)^2\,
  \BigLeftHook\frac{8}{3}\,\frac{1+\hat{\tau}^2}{1-\hat{\tau}}
  \ln^2\frac{Q^2}{m^2}}\nonumber\\[2ex]
  && + \BigLeftBrace\frac{1+\hat{\tau}^2}{1-\hat{\tau}}
  \BigLeftParen\frac{32}{3}\ln(1-\hat{\tau})
  -\frac{32}{3}\ln\hat{\tau} - \frac{80}{9}\BigRightParen
  -\frac{32}{3}(1-\hat{\tau})\BigRightBrace\,\ln\frac{Q^2}{m^2}
  \nonumber\\[2ex]
  && +\frac{1+\hat{\tau}^2}{1-\hat{\tau}}\BigLeftParen\frac{32}{3}
  \ln^2(1-\hat{\tau})-\frac{64}{3}\ln\hat{\tau}\,\ln(1-\hat{\tau})
  +\frac{28}{3}\ln^2\hat{\tau} - \frac{160}{9}\ln(1-\hat{\tau})
  \nonumber\\[2ex]
  && +\frac{160}{9}\ln\hat{\tau}-\frac{8}{3}\Li_2(1-\hat{\tau})
  -\frac{32}{3}\zeta(2) + \frac{448}{27}\BigRightParen
  -\frac{16}{3}\,(1-\hat{\tau})\cdot\nonumber\\[2ex]
  && \cdot\left(4\ln(1-\hat{\tau})-4\ln\hat{\tau}
  -\frac{19}{3}\right)-\frac{16}{3}\ln\hat{\tau}+\frac{8}{3}
  (1+\hat{\tau})\left(\Li_2(1-\hat{\tau})+\frac{1}{2}\ln^2\hat{\tau}\right)
  \BigRightHook.\nonumber\\[2ex]
  &&
\end{eqnarray}
If the parton cross section \eref{eq:parton_cross} is convoluted
by the parton densities in order to compute the hadronic cross sections,
as will be discussed in the next section, one approaches a singularity
at $\hat{\tau}=1$. In this case one cannot neglect the mass $m$ in the
denominator anymore. The resulting terms which even can go as a power of
the type $\ln^3 Q^2/m^2$ will be partially cancelled by similar terms arising
in the vertex correction \eref{eq:W_qbarq_V_2} (see \eref{eq:self_m_zero}).
They all
can be described by an expression proportional to a delta function which
has to be added to expression \eref{eq:W_qbarq_V_(1)}. This expression
reads as follows
\begin{eqnarray}
  \label{eq:W_qbarq_V_(2)}
  \lefteqn{\hat{W}_{q\bar{q}}^{V,(2)}\left(\hat{\tau},\frac{Q^2}{m^2}
  \right) =
  \left|V_{q_1\bar{q}_2}\right|^2
  \left\{\left(v_q^V\right)^2+\left(a_q^V\right)^2\right\}\,C_F T_f\,
  \left(\frac{\alpha_s}{4\pi}\right)^2\,\delta(1-\hat{\tau})\,
  \BigLeftHook4\ln^2\frac{Q^2}{m^2}}\nonumber\\[2ex]
  && -\frac{68}{3}\ln\frac{Q^2}{m^2} + \frac{32}{3}\zeta(3)
  -16\zeta(2) + \frac{454}{9}\BigRightHook.
\end{eqnarray}
Furthermore one has to replace in \eref{eq:W_qbarq_V_(1)} the singular
terms of the type $\ln^i(1-\hat{\tau})/(1-\hat{\tau})$ by
$(\ln^i(1-\hat{\tau})/(1-\hat{\tau}))_+$ with the definition
\begin{equation}
  \label{eq:def_distribution}
  \int_0^1 dx\,\left(\frac{\ln^i(1-x)}{1-x}\right)_+\,f(x) =
  \int_0^1dx\,\frac{\ln^i(1-x)}{1-x}\,\left(f(x)-f(1)\right).
\end{equation}
When the vector boson is radiated off the final state
(see Figs.~\ref{fig:W_qbarq_V_QbarQ}c,d)
one finds in the limit $Q^2 >> m^2$
\begin{eqnarray}
  \label{eq:W_qbarq_V_(3)}
  \lefteqn{\hat{W}_{q\bar{q}}^{V,(3)}\left(\hat{\tau},\frac{Q^2}{m^2}
  \right) =
  \left|V_{q_1\bar{q}_2}\right|^2
  \left\{\left(v_Q^V\right)^2+\left(a_Q^V\right)^2\right\}\,C_F T_f\,
  \left(\frac{\alpha_s}{4\pi}\right)^2\,\BigLeftHook
  (1+\hat{\tau})^2\BigLeftBrace}\nonumber\\[2ex]
  && -\frac{32}{3}\Li_2(-\hat{\tau}) - \frac{16}{3}\zeta(2)
  +\frac{8}{3}\ln^2\hat{\tau}-\frac{32}{3}\ln\hat{\tau}\,\ln(1+\hat{\tau})
  \BigRightBrace\nonumber\\[2ex]
  && + \frac{8}{3}(3 + 4\hat{\tau}+3\hat{\tau}^2)\ln\hat{\tau}
  + \frac{40}{3}(1-\hat{\tau}^2)\BigRightHook.
\end{eqnarray}
Finally in the case of $Z$-production one can also find an asymptotic
expression for the interference terms between
diagrams~\ref{fig:W_qbarq_V_QbarQ}a,b
and~\ref{fig:W_qbarq_V_QbarQ}c,d. It is given by \cite{Ham91,Gon92}
\begin{eqnarray}
  \label{eq:W_qbarq_V_(4)}
  \hat{W}_{q\bar{q}}^{Z,(4)}\left(\hat{\tau},\frac{Q^2}{m^2}\right) =
  a_q^Za_Q^Z\,C_F T_f\,\left(\frac{\alpha_s}{4\pi}\right)^2\,
  \BigLeftHook 16\frac{1 + \hat{\tau}^2}{1 - \hat{\tau}}\ln\hat{\tau}
  + 32\hat{\tau}\ln\hat{\tau} + 16(3-\hat{\tau})\BigRightHook.
  \hspace*{-1cm}\nonumber\\[2ex]
  &&
\end{eqnarray}
For the gluon-gluon fusion process \eref{eq:sub_gg_V_QbarQ} (see
Fig.~\ref{fig:W_gg_V_QbarQ}) one finds the following asymptotic expression
\begin{eqnarray}
  \label{eq:W_gg_V_(1)}
  \lefteqn{\hat{W}_{gg}^{V,(1)}\left(\hat{\tau},\frac{Q^2}{m^2}
  \right) =
  \left|V_{q_1\bar{q}_2}\right|^2
  \left\{\left(v_Q^V\right)^2+\left(a_Q^V\right)^2\right\}\,
  T_f^2\,\left(\frac{\alpha_s}{4\pi}\right)^2\,\BigLeftBrace
  \BigLeftHook-\{8(1+4\hat{\tau}+4\hat{\tau}^2)\cdot}
  \nonumber\\[2ex]
  && \ln\hat{\tau}+16(1-\hat{\tau})(1+3\hat{\tau})\}\ln^2\frac{Q^2}{m^2}
  +\BigLeftBrace -32(1+4\hat{\tau}+4\hat{\tau}^2)\BigLeftParen
  \Li_2(1-\hat{\tau}) +\nonumber\\[2ex]
  && \ln\hat{\tau}\,\ln(1-\hat{\tau})
  -\frac{1}{4}\ln^2\hat{\tau}\BigRightParen - 64(1-\hat{\tau})
  (1+3\hat{\tau})\ln(1-\hat{\tau}) +\nonumber\\[2ex]
  && 8(1+8\hat{\tau}-4\hat{\tau}^2)\ln\tau
  +4(1-\hat{\tau})(7+67\hat{\tau})\BigRightBrace\ln\frac{Q^2}{m^2}
  + 4(1+4\hat{\tau} +4\hat{\tau}^2)\cdot\nonumber\\[2ex]
  && \BigLeftParen 16\Li_3(1-\hat{\tau})
  -4\Li_2(1-\hat{\tau})\,\ln\hat{\tau} - 16\Li_2(1-\hat{\tau})\,
  \ln(1-\hat{\tau}) +\nonumber\\[2ex]
  && 4\ln^2\hat{\tau}\,\ln(1-\hat{\tau})
  -8\ln^2(1-\hat{\tau})\,\ln\hat{\tau}\BigRightParen + 4(1+\hat{\tau})
  \BigLeftParen 8\Li_2(-\hat{\tau}) +\nonumber\\[2ex]
  && 8\ln\hat{\tau}\,\ln(1+\hat{\tau})\BigRightParen
  +4(1+\hat{\tau})^2\BigLeftParen -16S_{1,2}(-\hat{\tau})
  -16\Li_2(-\hat{\tau})\ln(1+\hat{\tau})\nonumber\\[2ex]
  && - 8\zeta(2)\ln(1+\hat{\tau})
  +12\ln^2\hat{\tau}\,\ln(1+\hat{\tau}) - 8\ln^2(1+\hat{\tau})\,
  \ln\hat{\tau}\BigRightParen - 32(1+10\hat{\tau}+\nonumber\\[2ex]
  && 7\hat{\tau}^2)\,S_{1,2}(1-\hat{\tau})
  -32(1+2\hat{\tau}-\hat{\tau}^2)\Li_3(-\hat{\tau}) - 16
  (1+2\hat{\tau}-2\hat{\tau}^2)\zeta(3)\nonumber\\[2ex]
  && - 16(5+4\hat{\tau}-14\hat{\tau}^2)\Li_2(1-\hat{\tau})
  +32(2+4\hat{\tau}+\hat{\tau}^2)\Li_2(-\hat{\tau})\ln\hat{\tau}
  + 16\zeta(2)\cdot\nonumber\\[2ex]
  && (3+10\hat{\tau}+10\hat{\tau}^2)\ln\hat{\tau}
  +16(5+9\hat{\tau}-12\hat{\tau}^2)\zeta(2) -\frac{8}{3}(3+
  8\hat{\tau}+8\hat{\tau}^2)\ln^3\hat{\tau}\nonumber\\[2ex]
  && -8(3+7\hat{\tau}+4\hat{\tau}^2)\ln^2\hat{\tau}
  -64(1-\hat{\tau})(1+3\hat{\tau})\ln^2(1-\hat{\tau}) + 16
  (1+8\hat{\tau}-4\hat{\tau}^2)\cdot\nonumber\\[2ex]
  && \ln\hat{\tau}\,\ln(1-\hat{\tau})
  -4(23+64\hat{\tau}-105\hat{\tau}^2)\ln\hat{\tau} + 8(1-\hat{\tau})
  (7+67\hat{\tau})\ln(1-\hat{\tau})\nonumber\\[2ex]
  && -8(1-\hat{\tau})(16+49\hat{\tau})\BigRightHook\nonumber\\[2ex]
  && + \frac{N^2}{N^2-1}\BigLeftHook 4(1+\hat{\tau})^2\BigLeftBrace
  16 S_{1,2}(-\hat{\tau})+24\Li_3(-\hat{\tau})+16\zeta(3)\nonumber\\[2ex]
  && +\frac{16}{3}\Li_2(-\hat{\tau}) - 24\Li_2(-\hat{\tau})\ln\hat{\tau}
  +16\Li_2(-\hat{\tau})\ln(1+\hat{\tau}) + 8\zeta(2)\ln(1+\hat{\tau})
  \nonumber\\[2ex]
  && +\frac{8}{3}\zeta(2) - 12\ln^2\hat{\tau}\,\ln(1+\hat{\tau})
  +8\ln^2(1+\hat{\tau})\,\ln\hat{\tau} + \frac{16}{3}\ln\hat{\tau}\,
  \ln(1+\hat{\tau})\BigRightBrace\nonumber\\[2ex]
  && -32(1-\hat{\tau})^2S_{1,2}(1-\hat{\tau}) + \frac{8}{3}
  (-2+2\hat{\tau}+25\hat{\tau}^2)\ln^2\hat{\tau} - \frac{8}{3}
  (6+38\hat{\tau}+75\hat{\tau}^2)\ln\hat{\tau}\nonumber\\[2ex]
  && -\frac{1}{3}(1-\hat{\tau})(188+764\hat{\tau})\BigRightHook
  \BigRightBrace.
\end{eqnarray}
As we will see in the next section some of the above approximation turn
out to be very useful for $Z$-production accompanied with $b\bar{b}$
quarks because $Q^2 = M_Z^2 >> m_b^2$.\\
As has been discussed below \eref{eq:sub_gg_V_QbarQ} $W$-production has
to be treated in a different way as has been done above for $Z$- and
$\gamma$-production. Here one has to make a distinction between initial
and final state emission of the $W$-boson. In the case the $W$-boson
is radiated off from a light quark in the initial state, described
by the graphs in Figs.~\ref{fig:W_qbarq_V_QbarQ}a,b,
the DY structure function is given
by $\hat{W}_{q\bar{q}}^W$ in \eref{eq:W_ij_V}. However if the $W$-boson
is the decay product of the top or anti-top quark in the final state
like in Figs.~\ref{fig:W_qbarq_V_QbarQ}c,d
or Fig.~\ref{fig:W_gg_V_QbarQ} one has to resort to
different methods. In this paper we are only interested in the total
cross section. Hence we can follow the same procedure as is outlined in
\eref{eq:narrow_width_V}-\eref{eq:total_parton_W}. First we neglect
the graphs where the $W$ is emitted from the bottom quark because
the latter is far off-shell
and apply the narrow width approximation to the Breit-Wigner form of
the top quark in reactions \eref{eq:qbarq_top_1} and  \eref{eq:qbarq_top_2}.
This is a reasonable approach because the width of the top $\Gamma_t
=1.41\,{\rm GeV}$\footnote{$\Gamma_t$ is related to $m_t$ using the
formul\ae~in \cite{Den91}.} is much smaller than its mass $m_t = 174\,
{\rm GeV}$ \cite{Abe94}. Following the above procedure for $W$-production in
quark-antiquark annihilation the total cross section is
then given by
\begin{equation}
  \label{eq:cross_qbarq_W}
  \sigma_{q\bar{q}}^W = \sigma_{\rm tot}(q\bar{q}\rightarrow
  t\bar{t})\,B(t\rightarrow Wb),
\end{equation}
with $B(t\rightarrow Wb) \approx 1$ and $\sigma_{\rm tot}(q\bar{q}\rightarrow
t\bar{t})$ \cite{Jon78,*Glu78,*Bab78,*Geo78,*Com79,*Hag79,Lae92} is equal to
\begin{equation}
  \label{eq:cross_tot}
  \sigma_{\rm tot}(q\bar{q}\rightarrow t\bar{t}) = \frac{4\pi}{3}\,
  \alpha_s^2\frac{1}{N}\,C_F T_f\,\frac{1}{\hat{s}}
  \sqrt{1-\frac{4m^2}{\hat{s}}}\,\left(1+\frac{2m^2}{\hat{s}}\right).
\end{equation}
In the case of the gluon-gluon fusion process \eref{eq:sub_gg_V_QbarQ}
we proceed in the same way. Neglecting the emission of the $W$ from
the bottom quark and applying the narrow width approximation to
reactions \eref{eq:gg_top_1}, \eref{eq:gg_top_2} we get
\begin{equation}
  \label{eq:cross_gg_W}
  \sigma_{gg}^W = \sigma_{\rm tot}(gg\rightarrow t\bar{t})\,
  B(t\rightarrow Wb),
\end{equation}
where $\sigma_{\rm tot}(gg\rightarrow t\bar{t})$ is given by
\cite{Jon78,*Glu78,*Bab78,*Geo78,*Com79,*Hag79,Lae92}
\begin{eqnarray}
  \label{eq:cross_gg}
  \lefteqn{\sigma_{\rm tot}(gg\rightarrow t\bar{t}) = 4\pi\alpha_s^2
  \frac{1}{N}\,T_f^2\frac{1}{\hat{s}}\BigLeftHook\BigLeftBrace
  -\left( 1+\frac{4m^2}{\hat{s}}\right)\sqrt{1-\frac{4m^2}{\hat{s}}}}
  \nonumber\\[2ex]
  && + \left(1 + \frac{4m^2}{\hat{s}} - \frac{8m^4}{\hat{s}^2}\right)
  \ln y(\hat{s})\BigRightBrace + \frac{N^2}{N^2-1}\BigLeftBrace
  -\left(\frac{2}{3} + \frac{10}{3}\frac{m^2}{\hat{s}}\right)
  \sqrt{1-\frac{4m^2}{\hat{s}}}\nonumber\\[2ex]
  && +\frac{8m^4}{\hat{s}^2}\ln y(\hat{s})\BigRightBrace\BigRightHook,
\end{eqnarray}
with
\begin{equation}
  \label{eq:def_y}
  y(\hat{s}) = \frac{1+\sqrt{1-4m^2/\hat{s}}}
                    {1-\sqrt{1-4m^2/\hat{s}}}.
\end{equation}
\newpage
\section{Hadronic cross sections}
In this section we want to discuss the heavy flavor ($b$ and $t$)
contribution to the total cross section of $W$- and $Z$-production
at large hadron colliders. The energies and the colliders under study
are given by $\sqrt{s} = 0.63\,{\rm TeV}$ ($Sp\bar{p}S$, $p\bar{p}$),
$\sqrt{s} = 1.8\,{\rm TeV}$ (TEVATRON, $p\bar{p}$) and $\sqrt{s} = 16\,
{\rm TeV}$ (LHC, $pp$). Investigated will be the part of the total cross
section constituted by the heavy flavor contribution. We also want to
know how the latter, which is of order \alphastwo, compares with the
light parton contribution calculated in the same order of perturbation
theory which has been studied in the past (see \cite{Ham91,Nee92}).
Finally we want to study the validity of the
approximation for the total cross section of $H_1 + H_2 \rightarrow
Z + b + \bar{b}$ obtained from
the formul\ae~in \eref{eq:W_qbarq_V_(1)}-\eref{eq:W_gg_V_(1)}
which are calculated in the limit $M_Z^2 >> m_b^2$.\\
The hadronic cross section is related to the partonic cross
section \eref{eq:parton_cross} through the relation
\begin{eqnarray}
  \label{eq:hadron_cross}
  \lefteqn{\frac{d\sigma^V}{dQ^2} = \sum_{i,j}\,\int_0^1dy_1\,\int_0^1dy_2\,
  \int_0^1dz\,\delta(\tau-y_1y_2z)\,y_1y_2\,
  f_i^{H_1}(y_1,\mu^2)\,f_j^{H_2}(y_2,\mu^2)}\cdot\nonumber\\[2ex]
  && \frac{d\hat{\sigma}_{ij}^V}{dQ^2}(Q^2/y_1y_2s,Q^2,m_1^2,m_2^2,\mu^2),
\end{eqnarray}
where $f_i^H(y,\mu^2)$ denotes the density of parton $i$ inside
the hadron $H$ which depends aside from $y$ also on the factorization
(renormalization) scale $\mu$. Notice that for convenience we have put
the renormalization scale equal to the factorization scale.\\
In the case heavy flavor production is treated in lowest order, as we
do in this paper, $d\hat{\sigma}/dQ^2$ is independent of the
factorization scale (Born approximation). However since this approximation
is of order \alphastwo~it does depend on the renormalization scale. In
the case of light partons in the initial and final state one has to
perform mass factorization to $d\hat{\sigma}_{ij}/dQ^2$ in order to
remove the collinear divergences and this quantity has to be replaced
by the DY coefficient function which has been calculated up to order
\alphastwo~in \cite{Ham91,Nee92}. Therefore in addition to the
renormalization scale the latter also depends on the factorization
scale.\\
In our calculations we have chosen the \MS-scheme for the coupling
constant \alphas~as well as for the DY-coefficient function calculated up to
order \alphastwo~in \cite{Ham91}. For the parton
densities we have chosen the \MS-version of the set
MRS(H) \cite{Mar94} with $\Lambda_{\overline{\rm MS}}^{(4)} =
230\,{\rm MeV}$.
Furthermore we use the two-loop corrected running coupling constant
\alphas~with the QCD scale $\Lambda$ mentioned above. Since we only
consider top and bottom quark production we have put the number of light
flavors equal to four, i.e. $n_f = 4$. Finally we have set the
factorization (renormalization) scale $\mu^2 = Q^2$ where $Q^2 = M_V^2$.
For the electroweak parameters we have taken the following values: $M_Z
= 91.196\,{\rm GeV}$, $M_W = 80.24\,{\rm GeV}$, $G_F = 1.1667\ten{-5}
\,{\rm GeV}^{-2}$, $\sin^2\theta_W = 0.2258$ and $\sin^2\theta_C
= 0.0484$. The masses of the heavy flavors are given by $m_b = 5\,
{\rm GeV}$ and $m_t = 174\,{\rm GeV}$.\\
To calculate the total cross section for $Z$-production we have integrated
expression \eref{eq:hadron_cross} over $Q^2$ and used the narrow width
approximation \eref{eq:total_parton_Z}. In tables
\ref{tab:Z_630}-\ref{tab:Z_16000}
we have listed the various contributions coming from the partonic
subprocesses discussed in the previous section and compared them with
the light parton cross section corrected up to order \alphastwo. The
tables reveal that at smaller energies, i.e. $\sqrt{s} = 0.63\,{\rm TeV}$
the total heavy flavor cross section is dominated by the vertex
correction \eref{eq:W_qbarq_Z_2} given by Fig.~\ref{fig:W_qbarq_Z_2} and
the subprocess $q+\bar{q}\rightarrow Z+b+\bar{b}$ \eref{eq:W_ij_V}
depicted in Fig.~\ref{fig:W_qbarq_V_QbarQ}.
As we will show later on the importance of the last process is wholly due
to initial state radiation (Figs.~\ref{fig:W_qbarq_V_QbarQ}a,b).
At larger energies like
$\sqrt{s} = 1.8\,{\rm TeV}$ also the process $g+g\rightarrow Z+b+\bar{b}$
\eref{eq:W_ij_V} (Fig.~\ref{fig:W_gg_V_QbarQ}) becomes important. The
latter even overwhelms the other reactions when $\sqrt{s} = 16\,
{\rm TeV}$ (LHC). This can be traced back  to the gluon density which steeply
rises at very small $\tau = M_V^2/s$.
Finally we observe that the processes with top
quarks in the final state are completely unimportant which is due
to the limited phase space available even for energies as large as $\sqrt{s} =
16\,{\rm TeV}$ (LHC).\\
The reason that the virtual
correction in Fig.~\ref{fig:W_qbarq_Z_2} plays an important role can
be inferred from the expression in \eref{eq:W_qbarq_Z_2}. Here the first
term $G_1$ only contributes in the case of the bottom loop ($M_Z^2 >
4m_b^2$) whereas the second term $G_2$ only contributes for the top
loop ($M_Z^2 < 4m_t^2$). In \cite{Gon92} it has been shown that for
$Q^2 >> m^2$ the function $G_1$ vanishes like
\begin{equation}
  \label{eq:limit_G1}
  G_1\left(\frac{m^2}{Q^2}\right) \sim {\cal O}\left(\frac{m^2}{Q^2}
  \right),
\end{equation}
whereas the function $G_2$ has the following asymptotic behavior for
$m^2 >> Q^2$
\begin{equation}
  \label{eq:limit_G2}
  G_2\left(\frac{m^2}{Q^2}\right) \sim -3\ln\frac{Q^2}{m^2}
  -2\zeta(2) + \frac{21}{2} + {\cal O}\left(\frac{Q^2}{m^2}\right),
\end{equation}
which means that this correction is dominated by the top-loop contribution.\\
In tables~\ref{tab:Z_630}-\ref{tab:Z_16000} we have
also listed the results coming from the light parton contribution
calculated in \cite{Ham91}. We observe that the heavy flavor
part of the DY cross section is very small and it amounts to 0.2\%
($\sqrt{s} = 0.63\,{\rm TeV})$, 0.3\% ($\sqrt{s} = 1.8\,{\rm TeV}$),
1\% ($\sqrt{s} = 16\,{\rm TeV})$ of the light quark contribution.
Even if we compare the heavy flavor part, which is an order \alphastwo~
correction, with the corresponding light parton contribution in the
same order of perturbation theory one discovers that it is very small
except when the energy gets very large.
It amounts to 5\% ($\sqrt{s} = 0.63\,{\rm TeV}$), 11\%
($\sqrt{s} = 1.8\,{\rm TeV}$), and 770\% (absolute value)
($\sqrt{s} = 16\,{\rm TeV}$) of
the \alphastwo~correction to the cross section which is due to light partons.
This means that only at LHC energies heavy flavor production is more
important than the light parton subprocesses contributing in order
\alphastwo.\\
For the $W$-cross section we
proceed in the same way as for $Z$-production except that here we also have
to make the narrow width approximation for the top quark. This is
needed when the $W$ is radiated off the top-quark in the final state
exhibited by the graphs in Figs.~\ref{fig:W_qbarq_V_QbarQ}c,d and
\ref{fig:W_gg_V_QbarQ}. Furthermore the graphs containing the triangle
heavy flavor loop like Figs.~\ref{fig:W_qbarq_Z_2}, \ref{fig:W_qbarq_Z_1},
and \ref{fig:W_qg_Z} do not contribute. The results are given in
tables~\ref{tab:W_630}-\ref{tab:W_16000}. Like in the case of
$Z$-production the
process $q_1+\bar{q}_2\rightarrow W+b+\bar{b}$ \eref{eq:W_ij_V}
(Figs.~\ref{fig:W_qbarq_V_QbarQ}a,b) is dominant at lower energies
although also the vertex correction \eref{eq:W_qbarq_V_2} contributes
a little bit. When the energy gets larger like in the case of LHC also
the process $g+g\rightarrow t+\bar{t}$ with $t\rightarrow W^+b$ and
$\bar{t}\rightarrow W^-\bar{b}$ \eref{eq:cross_gg_W} (Fig.
\ref{fig:W_gg_V_QbarQ}) becomes important due to the gluon density
which becomes very large at small $\tau = M_W^2/s$. As in $Z$-production
the heavy flavors give a small contribution to the $W$-cross section.
The latter is of the same magnitude as in the $Z$-cross section and it
amounts to 0.1\% ($\sqrt{s} = 0.63\,{\rm TeV}$), 0.1\% ($\sqrt{s}=
1.8\,{\rm TeV}$), and 0.2\% ($\sqrt{s} = 16\,{\rm TeV}$) of the light
parton contribution. If we make a comparison with the order \alphastwo~
part of the light parton contribution these numbers become 2\%,
6\%, and 90\% respectively which are however smaller than in the case
of $Z$-production.\\
Besides vector boson production we have also studied the heavy flavor
part of the DY cross section $d\sigma/dQ^2$ at small $Q^2$
($\sqrt{Q^2} < 60\,{\rm GeV}$) where the virtual photon dominates the
cross section. In Fig.~\ref{fig:photo_prod} we have plotted the ratio $R$
\begin{equation}
  \label{eq:ratio_R}
  R(Q^2) = \frac{\displaystyle\frac{d\sigma}{dQ^2}(u,d,s,c,g)
                 +\frac{d\sigma}{dQ^2}(t,b)}
                {\displaystyle\frac{d\sigma}{dQ^2}(u,d,s,c,g)},
\end{equation}
at three different energies, i.e. $\sqrt{s} = 0.63, 1.8, 16\,{\rm TeV}$.
Like in the case of $Z$- and $W$-boson production the contribution of the
heavy flavors to the DY cross section is very small. When the energy
increases it grows from 0.1\% to 0.5\% of the part constituted by the
light parton contributions.\\
Using the MRS(H) \cite{Mar94} parton densities we also
calculate the cross sections for $Z$- and $W$-production where the lepton
pair into which the vector boson decays is measured. The results are
presented in table~\ref{tab:branch_630} ($Sp\bar{p}S$, $\sqrt{S}=0.63\,
{\rm TeV}$)
and table~\ref{tab:branch_1800} (TEVATRON, $\sqrt{s}=1.8\,{\rm TeV}$).
They are obtained by multiplying the total cross sections $\sigma^Z$
and $\sigma^W$ in tables \ref{tab:Z_630} to \ref{tab:W_16000}
by the branching ratios $B(Z\rightarrow e^+e^-) = 3.35\ten{-2}$
and $B(W\rightarrow e\nu_e) = 0.109$ respectively. Furthermore we have also
listed the experimental data obtained from the groups UA1 \cite{Alb91},
UA2 \cite{Ali91} ($Sp\bar{p}S$) and CDF \cite{Abe92} (TEVATRON). As is
expected from the previous discussion the contribution from the heavy
flavors $b$ and $t$ to the cross section is extremely small when
compared with the light parton part.\\
Before concluding this section we have also studied the cross sections
obtained from \eref{eq:W_qbarq_V_(1)}-\eref{eq:W_gg_V_(1)} which are
derived in the limit $Q^2 >> m^2$. In practice these formul\ae~are
only applicable to the reaction $H_1 + H_2 \rightarrow Z+b+\bar{b}$ where
$M_Z^2 >> m_b^2$. This inequality does not apply to top production
also when the $Z$ is replaced by the $W$.
In the case a photon appears in the intermediate
state the above formulae are also not very useful because in the region
$Q^2 >> m^2$ the production rate is too low.
In table~\ref{tab:comparison} we have compared the results
obtained from the exact cross section represented by \eref{eq:W_qbarq_V_2},
\eref{eq:W_ij_V} with those predicted by the asymptotic formul\ae~in
\eref{eq:W_qbarq_V_(1)}-\eref{eq:W_gg_V_(1)}. First the table shows that the
whole contribution to the cross section of the subprocess
$q+\bar{q}\rightarrow Z+b+\bar{b}$ is given by the initial state
radiation of the $Z$-boson as
depicted in Figs.~\ref{fig:W_qbarq_V_QbarQ}a,b which also includes the
virtual contribution in Fig.~\ref{fig:W_qbarq_V_2}. This is due to the large
logarithmic terms $\ln^k(Q^2/m^2)$ which
show up in \eref{eq:W_qbarq_V_(1)} and \eref{eq:W_qbarq_V_(2)}. Furthermore
the approximations to the initial state radiation process
work rather well in particularly when the energy increases. The difference
between the exact \eref{eq:W_qbarq_V_2}, \eref{eq:W_ij_V} and the approximate
cross sections \eref{eq:W_qbarq_V_(1)}, \eref{eq:W_qbarq_V_(2)} amounts to
30\% ($\sqrt{s} = 0.63\,{\rm TeV}$), 17\% ($\sqrt{s} = 1.8\,{\rm TeV}$) and
16\% ($\sqrt{s} = 16\,{\rm TeV}$) of the exact value. The main reason for the
difference is that the cubic logarithmic term $\ln^3(Q^2/m^2)$, arising
in the exact expressions \eref{eq:W_qbarq_V_2}, \eref{eq:W_ij_V} in the limit
$Q^2 >> m^2$, will only cancel when $m^2$ is taken to be really small with
respect to $Q^2$. This is apparently not the case for $Q^2 = M_Z^2$ and
$m^2 = m_b^2$. Therefore the sum of the two exact expressions is not quite
equal to the sum of the approximations \eref{eq:W_qbarq_V_(1)} and
\eref{eq:W_qbarq_V_(2)} which behaves asymptotically like $\ln^2(Q^2/m^2)$.
In practice the asymptotic limit is only reached for $m^2 < 1 ({\rm GeV}/c)^2$
which is an order of magnitude less than $m^2 = (5\,{\rm GeV}/c^2)^2$.
A similar problem shows up in the contribution
originating from the interference between initial state
(Figs.~\ref{fig:W_qbarq_V_QbarQ}a,b) and
final state (Figs.~\ref{fig:W_qbarq_V_QbarQ}c,d) radiation of the vector boson.
In spite of the fact that here no large logarithms of the type
$\ln^k(Q^2/m^2)$ appear in the final expression \eref{eq:W_qbarq_V_(4)} the
various integrals contributing to the interference term contain these type
of logarithms so that in this case one observes an incomplete cancellation too.
{}From table~\ref{tab:comparison} we infer that the approximation to the
interference term is not so bad when $\sqrt{s} = 0.63\,{\rm TeV}$ but it
becomes worse at higher energies. Apart from the incomplete cancellation
mentioned
above this is also due to the fact that the quality of the approximation
depends on the values for $\hat{s} = y_1 y_2 s$ in the partonic
cross section $d\hat{\sigma}/dQ^2$ \eref{eq:hadron_cross}.
When the energy $\sqrt{s}$ increases it may
happen that the product of the parton densities in \eref{eq:hadron_cross}
probe the $\hat{s}$-region where the approximation fails.
Fortunately the interference term gives a negligible contribution to the
cross section of $q+\bar{q}\rightarrow Z+b+\bar{b}$. The latter also holds
for the final state radiation of the $Z$-boson as depicted by the graphs
in Figs.~\ref{fig:W_qbarq_V_QbarQ}c,d. However here the approximation to the
the cross section \eref{eq:W_ij_V} as
given by \eref{eq:W_qbarq_V_(3)} becomes excellent which is independent
of the energies under consideration.
The most simple explanation for this phenomenon is that here
large logarithms of the type $\ln^k(Q^2/m^2)$
neither appear in the final result
\eref{eq:W_qbarq_V_(3)} nor in the separate integrals contributing to this
expression. Therefore in this case no cancellation of large logarithmic terms
has to occur. Finally table~\ref{tab:comparison} reveals that the approximation
to the cross section of the gluon-gluon fusion process $g+g\rightarrow Z+b+
\bar{b}$ (Fig.~\ref{fig:W_gg_V_QbarQ}) turns out to be very good for
all energies
under consideration. Although the approximate cross section
\eref{eq:W_gg_V_(1)}
behaves quadratically in $\ln(Q^2/m^2)$ it is closer to the exact result as
discovered for the initial state radiation of the $Z$-boson in
$q+\bar{q}\rightarrow Z+b+\bar{b}$. This is mainly due to the fact that in the
former process no cancellation of leading logarithmic terms occur.\\
In general we can conclude that irrespective of the energies considered the
approximations work rather well for $Z$-production accompanied by
bottom quarks although this statement is more valid for $g+g\rightarrow
Z+b+\bar{b}$ (Fig.~\ref{fig:W_gg_V_QbarQ}) than for $q+\bar{q}\rightarrow
Z+b+\bar{b}$
(Fig.~\ref{fig:W_qbarq_V_QbarQ}). Since the large logarithms $\ln^k(Q^2/m^2)$
dominate the cross sections for $\sqrt{s}> 1.8 \,{\rm TeV}$ the bottom
can be treated as a light quark. Further we have also studied the limit
$Q^2 >> m^2$ for the charmed quark cross sections. In this case the
approximations \eref{eq:W_qbarq_V_(1)}-\eref{eq:W_gg_V_(1)} are even better
than for bottom production. This even holds for $\sqrt{s} = 0.63\,{\rm TeV}$.
The logarithms of the type $\ln(Q^2/m^2)$ dominate the charm and bottom cross
sections and they give rise to large corrections. Therefore they have to be
removed by mass factorization and subsequently to be absorbed, after
resummation via the renormalization group equations, into the charm and bottom
densities in the hadron.\\
Summarizing the content of this paper we have computed all order
\alphastwo~contributions to the DY cross section which can be
attributed to heavy flavors. All virtual as well as radiative processes
have been considered. In this way we have extended the work done in
\cite{Dic86,Gon92} where only the contributions characteristic for
$Z$-production have been considered.\\
{}From the results obtained in this work one can conclude that the
contributions of the heavy flavors $b$ and $t$ to the DY cross section
in particularly to vector boson production are very small. They are on the one
percent level in the case of $Z$-production provided the energy is
very large which will only happen when the
LHC is put into operation. Furthermore we have shown that for $\sqrt{s}>1.8\,
{\rm TeV}$ the cross sections
\eref{eq:W_qbarq_V_(1)}-\eref{eq:W_gg_V_(1)} derived in the limit $Q^2 >> m^2$
can be applied to $Z$-$b\bar{b}$-production.
This means that the bottom quark can be treated as a light flavor at
large collider energies.
The heavy flavor cross sections will only become
observable if the vector boson as well as the heavy quarks are
detected. This will happen for the LHC where the process $p+p\rightarrow
Z+b+\bar{b}$ constitutes an important background for Higgs production
\cite{Gun90}.
\newpage
\appendix
\section{}
In this appendix we present the two-loop vertex correction defined by
$F(Q^2,m^2)$ in \eref{eq:W_qbarq_V_2}. It contains the gluon self energy
contribution with the heavy flavors appearing in the subloop
(Fig.~\ref{fig:W_qbarq_V_2}). The vertex correction reads
\begin{eqnarray}
  \label{eq:F_qbarq_V_2}
  \lefteqn{F(Q^2,m^2) = -\BigLeftParen
    \frac{440}{9}\frac{m^2}{Q^2} + \frac{530}{27}\BigRightParen
    \ln\frac{Q^2}{m^2} + x \BigLeftParen \frac{184}{9}
    \frac{m^2}{Q^2} + \frac{76}{9}\BigRightParen\cdot}\nonumber\\[2ex]
  && \BigLeftParen2\Li_2\BigLeftParen\frac{x-1}{x+1}\BigRightParen
    -2\zeta(2) + \frac{1}{2}\ln^2\frac{x-1}{x+1}\BigRightParen
    +\BigLeftParen 16\,\frac{m^4}{Q^4}-\frac{8}{3}\BigRightParen
    \cdot\nonumber\\[2ex]
  && \BigLeftParen-2\Li_3\BigLeftParen\frac{x-1}{x+1}\BigRightParen
    +2\zeta(3) - \frac{1}{6}\ln^3\frac{x-1}{x+1} + 2\zeta(2)
    \ln\frac{x-1}{x+1}\BigRightParen + \frac{952}{9}\frac{m^2}{Q^2}
    \nonumber\\[2ex]
  && +\frac{3355}{81},
\end{eqnarray}
where
\begin{equation}
  x = \sqrt{1+4\frac{m^2}{Q^2}}.
\end{equation}
Further the following asymptotic expansions turn out to be useful.
In the limit $m^2 << Q^2$ one obtains the expansion
\begin{eqnarray}
  \label{eq:self_m_zero}
  F(Q^2,m^2)&\begin{array}[t]{c}=\\{\scriptstyle m\rightarrow 0}\end{array}&
  -\frac{4}{9}\,\ln^3\frac{Q^2}{m^2} + \frac{38}{9}\,\ln^2\frac{Q^2}{m^2}
  +\BigLeftParen\frac{16}{3}\zeta(2)-\frac{530}{27}\BigRightParen\,
  \ln\frac{Q^2}{m^2} \nonumber\\[2ex]
  && + \frac{3355}{81} - \frac{152}{9}\zeta(2) -\frac{16}{3}\zeta(3).
\end{eqnarray}
In the case that $m^2 >> Q^2$ the vertex correction behaves like
\begin{equation}
  \label{eq:self_inf}
  F(Q^2,m^2)\begin{array}[t]{c}=\\{\scriptstyle m\rightarrow\infty}\end{array}
  \frac{Q^2}{m^2}\,\BigLeftParen\frac{176}{225} -
  \frac{8}{45}\,\ln\frac{Q^2}{m^2}\BigRightParen,
\end{equation}
which shows that heavy flavors decouple from $F(Q^2,m^2)$ when the
quark mass gets infinite.

\newpage
\begin{mytable}{|l|c|c|}{tab:Z_630}{Contributions to the total cross
section for $Z$-production at $\sqrt{s} = 0.63\,{\rm TeV}$
($\alpha_s(M_Z) = 0.107 $).}
\hline
subprocess & equation & $\sigma^Z$ (nb) \\ \hline
$q+\bar{q}\rightarrow Z$ & \eref{eq:W_qbarq_Z_2} & $3.13\ten{-3}$ \\ \hline
$q+\bar{q}\rightarrow Z$ & \eref{eq:W_qbarq_V_2} & $1.40\ten{-4}$ \\ \hline
$q+\bar{q}\rightarrow Z+g$ & \eref{eq:W_qbarq_Z_1} & $1.13\ten{-4}$ \\ \hline
$g+q(\bar{q})\rightarrow Z+q(\bar{q})$ & \eref{eq:W_qg_Z} & $-1.77\ten{-5}$
  \\ \hline
$q+\bar{q}\rightarrow Z+b+\bar{b}$ & \eref{eq:W_ij_V} & $1.47\ten{-3}$
  \\ \hline
$q+\bar{q}\rightarrow Z+t+\bar{t}$ & \eref{eq:W_ij_V} & $1.35\ten{-11}$
  \\ \hline
$g+g\rightarrow Z+b+\bar{b}$ & \eref{eq:W_ij_V} & $6.39\ten{-5}$ \\ \hline
$g+g\rightarrow Z+t+\bar{t}$ & \eref{eq:W_ij_V} & $1.25\ten{-16}$
  \\ \hline \hline
\multicolumn{2}{|l}{$\sigma^Z(b,t)$} & $4.90\ten{-3}$ \\ \hline
\multicolumn{3}{|l|}{
  $\sigma^Z(u,d,s,c,g)= 1.54\,{\rm (Born)} + 0.41\,(\calO(\alpha_s))
  + 0.10\,(\calO(\alpha_s)) = 2.05$} \\ \hline
\end{mytable}
\begin{mytable}{|l|c|c|}{tab:Z_1800}{Contributions to the total cross
section for $Z$-production at $\sqrt{s} = 1.8\,{\rm TeV}$
($\alpha_s(M_Z) = 0.107$).}
\hline
subprocess & equation & $\sigma^Z$ (nb) \\ \hline
$q+\bar{q}\rightarrow Z$ & \eref{eq:W_qbarq_Z_2} & $5.33\ten{-3}$ \\ \hline
$q+\bar{q}\rightarrow Z$ & \eref{eq:W_qbarq_V_2} & $4.86\ten{-4}$ \\ \hline
$q+\bar{q}\rightarrow Z+g$ & \eref{eq:W_qbarq_Z_1} & $3.67\ten{-4}$ \\ \hline
$g+q(\bar{q})\rightarrow Z+q(\bar{q})$ & \eref{eq:W_qg_Z} & $-1.61\ten{-4}$
  \\ \hline
$q+\bar{q}\rightarrow Z+b+\bar{b}$ & \eref{eq:W_ij_V} & $8.57\ten{-3}$
  \\ \hline
$q+\bar{q}\rightarrow Z+t+\bar{t}$ & \eref{eq:W_ij_V} & $3.58\ten{-6}$
  \\ \hline
$g+g\rightarrow Z+b+\bar{b}$ & \eref{eq:W_ij_V} & $4.71\ten{-3}$ \\ \hline
$g+g\rightarrow Z+t+\bar{t}$ & \eref{eq:W_ij_V} & $5.50\ten{-8}$
  \\ \hline \hline
\multicolumn{2}{|l}{$\sigma^Z(b,t)$} & $1.93\ten{-2}$ \\ \hline
\multicolumn{3}{|l|}{
  $\sigma^Z(u,d,s,c,g)= 5.34\,{\rm (Born)} + 1.05\,(\calO(\alpha_s))
  + 0.17\,(\calO(\alpha_s^2)) = 6.56$} \\ \hline
\end{mytable}
\begin{mytable}{|l|c|c|}{tab:Z_16000}{Contributions to the total cross
section for $Z$-production at $\sqrt{s} = 16\,{\rm TeV}$
($\alpha_s(M_Z) = 0.107$).}
\hline
subprocess & equation & $\sigma^Z$ (nb) \\ \hline
$q+\bar{q}\rightarrow Z$ & \eref{eq:W_qbarq_Z_2} & $1.21\ten{-2}$ \\ \hline
$q+\bar{q}\rightarrow Z$ & \eref{eq:W_qbarq_V_2} & $5.02\ten{-3}$ \\ \hline
$q+\bar{q}\rightarrow Z+g$ & \eref{eq:W_qbarq_Z_1} & $8.06\ten{-4}$ \\ \hline
$g+q(\bar{q})\rightarrow Z+q(\bar{q})$ & \eref{eq:W_qg_Z} & $-1.57\ten{-3}$
  \\ \hline
$q+\bar{q}\rightarrow Z+b+\bar{b}$ & \eref{eq:W_ij_V} & $9.28\ten{-2}$
  \\ \hline
$q+\bar{q}\rightarrow Z+t+\bar{t}$ & \eref{eq:W_ij_V} & $3.17\ten{-4}$
  \\ \hline
$g+g\rightarrow Z+b+\bar{b}$ & \eref{eq:W_ij_V} & $5.85\ten{-1}$ \\ \hline
$g+g\rightarrow Z+t+\bar{t}$ & \eref{eq:W_ij_V} & $9.23\ten{-4}$
  \\ \hline \hline
\multicolumn{2}{|l}{$\sigma^Z(b,t)$} & $6.95\ten{-1}$ \\ \hline
\multicolumn{3}{|l|}{
  $\sigma^Z(u,d,s,c,g)= 55.2\,{\rm (Born)} + 7.45\,(\calO(\alpha_s))
  - 0.09\,(\calO(\alpha_s^2)) = 62.6$} \\ \hline
\end{mytable}
\begin{mytable}{|l|c|c|}{tab:W_630}
{Contributions to the total cross section for $W^++W^-$-production at
$\sqrt{s}=0.63\,{\rm TeV}$ ($\alpha_s(M_W) = 0.109$).}
\hline
subprocess & equation & $\sigma^W$ (nb) \\ \hline
$q_1+\bar{q}_2\rightarrow W$ & \eref{eq:W_qbarq_V_2} & $1.32\ten{-3}$
  \\ \hline
$q_1+\bar{q}_2\rightarrow W+b+\bar{b}$ & \eref{eq:W_ij_V} & $4.11\ten{-3}$
  \\ \hline
$q_1+\bar{q}_2\rightarrow W+t+\bar{t}$ & \eref{eq:W_ij_V} & $1.94\ten{-11}$
  \\ \hline
\begin{tabular}{rcl}
  $q + \bar{q}$ &$\rightarrow$& $t(\bar{t}) + \bar{t}(t)$ \\
  && $\lfloor\hspace*{-2pt}\raisebox{-5.8pt}{$\rightarrow
    W^+(W^-) + b(\bar{b})$}$
\end{tabular}
& \eref{eq:cross_qbarq_W} & $1.43\ten{-6}$ \\ \hline
\begin{tabular}{rcl}
  $g + g$ &$\rightarrow$& $t(\bar{t}) + \bar{t}(t)$ \\
  && $\lfloor\hspace*{-2pt}\raisebox{-5.8pt}{$\rightarrow
    W^+(W^-) + b(\bar{b})$}$
\end{tabular}
& \eref{eq:cross_gg_W} & $1.52\ten{-9}$ \\ \hline
\multicolumn{2}{|l}{$\sigma^W(b,t)$} & $5.43\ten{-3}$ \\ \hline
\multicolumn{3}{|l|}{
  $\sigma^W(u,d,s,c,g)= 5.09\,{\rm (Born)} + 1.34\,(\calO(\alpha_s))
  + 0.33\,(\calO(\alpha_s^2)) = 6.76$} \\ \hline
\end{mytable}
\begin{mytable}{|l|c|c|}{tab:W_1800}
{Contributions to the total cross section for $W^++W^-$-production at
$\sqrt{s}=1.8\,{\rm TeV}$ ($\alpha_s(M_W) = 0.109$).}
\hline
subprocess & equation & $\sigma^W$ (nb) \\ \hline
$q_1+\bar{q}_2\rightarrow W$ & \eref{eq:W_qbarq_V_2} & $4.67\ten{-3}$
  \\ \hline
$q_1+\bar{q}_2\rightarrow W+b+\bar{b}$ & \eref{eq:W_ij_V} & $2.51\ten{-2}$
  \\ \hline
$q_1+\bar{q}_2\rightarrow W+t+\bar{t}$ & \eref{eq:W_ij_V} & $6.36\ten{-6}$
  \\ \hline
\begin{tabular}{rcl}
  $q + \bar{q}$ &$\rightarrow$& $t(\bar{t}) + \bar{t}(t)$ \\
  && $\lfloor\hspace*{-2pt}\raisebox{-5.8pt}{$\rightarrow
    W^+(W^-) + b(\bar{b})$}$
\end{tabular}
& \eref{eq:cross_qbarq_W} & $1.10\ten{-3}$ \\ \hline
\begin{tabular}{rcl}
  $g + g$ &$\rightarrow$& $t(\bar{t}) + \bar{t}(t)$ \\
  && $\lfloor\hspace*{-2pt}\raisebox{-5.8pt}{$\rightarrow
    W^+(W^-) + b(\bar{b})$}$
\end{tabular}
& \eref{eq:cross_gg_W} & $1.28\ten{-4}$ \\ \hline
\multicolumn{2}{|l}{$\sigma^W(b,t)$} & $3.10\ten{-2}$ \\ \hline
\multicolumn{3}{|l|}{
  $\sigma^W(u,d,s,c,g)= 18.0\,{\rm (Born)} + 3.47\,(\calO(\alpha_s))
 + 0.50\,(\calO(\alpha_s^2)) = 22.0$} \\ \hline
\end{mytable}
\begin{mytable}{|l|c|c|}{tab:W_16000}
{Contributions to the total cross section for $W^++W^-$-production at
$\sqrt{s}=16\,{\rm TeV}$ ($\alpha_s(M_W) = 0.109$).}
\hline
subprocess & equation & $\sigma^W$ (nb) \\ \hline
$q_1+\bar{q}_2\rightarrow W$ & \eref{eq:W_qbarq_V_2} & $4.79\ten{-2}$
  \\ \hline
$q_1+\bar{q}_2\rightarrow W+b+\bar{b}$ & \eref{eq:W_ij_V} & $2.76\ten{-1}$
  \\ \hline
$q_1+\bar{q}_2\rightarrow W+t+\bar{t}$ & \eref{eq:W_ij_V} & $6.26\ten{-4}$
  \\ \hline
\begin{tabular}{rcl}
  $q + \bar{q}$ &$\rightarrow$& $t(\bar{t}) + \bar{t}(t)$ \\
  && $\lfloor\hspace*{-2pt}\raisebox{-5.8pt}{$\rightarrow
    W^+(W^-) + b(\bar{b})$}$
\end{tabular}
& \eref{eq:cross_qbarq_W} & $2.00\ten{-2}$ \\ \hline
\begin{tabular}{rcl}
  $g + g$ &$\rightarrow$& $t(\bar{t}) + \bar{t}(t)$ \\
  && $\lfloor\hspace*{-2pt}\raisebox{-5.8pt}{$\rightarrow
    W^+(W^-) + b(\bar{b})$}$
\end{tabular}
& \eref{eq:cross_gg_W} & $1.77\ten{-1}$ \\ \hline
\multicolumn{2}{|l}{$\sigma^W(b,t)$} & $5.22\ten{-1}$ \\ \hline
\multicolumn{3}{|l|}{
  $\sigma^W(u,d,s,c,g)= 185\,{\rm (Born)} + 24.8\,(\calO(\alpha_s))
  - 0.6\,(\calO(\alpha_s^2)) = 209$} \\ \hline
\end{mytable}
\begin{mytable}{|c|c|c|}{tab:branch_630}
{$B_Z\,\sigma^Z$ and $B_W\,\sigma^W$ for the
$Sp\bar{p}S$ \cite{Alb91,Ali91} with $B_Z\equiv B(Z\rightarrow e^+e^-)
= 3.35\ten{-2}$ and $B_W\equiv B(W\rightarrow e\nu_e) = 0.109$,
$\sqrt{s} = 0.63\,{\rm TeV}$.}
\hline
 & $B_Z\,\sigma^Z$ (pb) & $B_W\,\sigma^W$ (pb) \\ \hline
 UA1 & $58.6\pm 7.8\pm 8.4$ & $609\pm 41\pm 94$ \\ \hline
 UA2 & $65.6\pm 4.0\pm 3.8$ & $682\pm 12\pm 40$ \\ \hline \hline
 $B_V\,\sigma^V(u,d,s,c,g)$ & 68.7 & 737 \\ \hline
 $B_V\,\sigma^V(b,t)$ & 0.164 & 0.592 \\ \hline
\end{mytable}
\begin{mytable}{|c|c|c|}{tab:branch_1800}
{$B_Z\,\sigma^Z$ and $B_W\,\sigma^W$ for the
TEVATRON \cite{Abe92} with $B_Z\equiv B(Z\rightarrow e^+e^-)
= 3.35\ten{-2}$ and $B_W\equiv B(W\rightarrow e\nu_e) = 0.109$,
$\sqrt{s} = 1.8\,{\rm TeV}$.}
\hline
 & $B_Z\,\sigma^Z$ (nb) & $B_W\,\sigma^W$ (nb) \\ \hline
 CDF & $0.214\pm 0.011\pm 0.020$ & $2.20\pm 0.04\pm 0.20$ \\ \hline \hline
 $B_V\,\sigma^V(u,d,s,c,g)$ & 0.220 & 2.40 \\ \hline
 $B_V\,\sigma^V(b,t)$ & $6.47\ten{-4}$ & $3.38\ten{-3}$ \\ \hline
\end{mytable}
\begin{mytable}{|l|cc|cc|}{tab:comparison}
{Comparison of the exact versus approximate cross sections for $Z$- and
$b\bar{b}$-production ($\alpha_s(M_Z) = 0.107$).}
\hline
\multicolumn{5}{|c|}{$\sqrt{s} = 0.63\,{\rm TeV}$} \\
 & $\sigma_{\rm exact}^Z$ (nb) & & $\sigma_{\rm app.}^Z$ (nb) & \\ \hline
$q+\bar{q}\rightarrow Z+b+\bar{b}$ & $1.58\ten{-3}$ & \eref{eq:W_qbarq_V_2},
  \eref{eq:W_ij_V} & $1.11\ten{-3}$ & \eref{eq:W_qbarq_V_(1)},
  \eref{eq:W_qbarq_V_(2)} \\ \cline{2-5}
 & $2.41\ten{-6}$ & \eref{eq:W_ij_V} & $2.48\ten{-6}$ &
  \eref{eq:W_qbarq_V_(3)} \\ \cline{2-5}
 & $-9.09\ten{-6}$ & \eref{eq:W_ij_V} & $-1.26\ten{-5}$ &
  \eref{eq:W_qbarq_V_(4)} \\ \hline
$g+g\rightarrow Z+b+\bar{b}$ & $6.39\ten{-5}$ & \eref{eq:W_ij_V} &
 $5.80\ten{-5}$ & \eref{eq:W_gg_V_(1)}
 \\ \hline \hline
\multicolumn{5}{|c|}{$\sqrt{s} = 1.8\,{\rm TeV}$} \\
 & $\sigma_{\rm exact}^Z$ (nb) & & $\sigma_{\rm app.}^Z$ (nb) & \\ \hline
$q+\bar{q}\rightarrow Z+b+\bar{b}$ & $8.88\ten{-3}$ & \eref{eq:W_qbarq_V_2},
  \eref{eq:W_ij_V} & $7.40\ten{-3}$ & \eref{eq:W_qbarq_V_(1)},
  \eref{eq:W_qbarq_V_(2)} \\ \cline{2-5}
 & $4.31\ten{-5}$ & \eref{eq:W_ij_V} & $4.35\ten{-5}$ &
  \eref{eq:W_qbarq_V_(3)} \\ \cline{2-5}
 & $-1.40\ten{-5}$ & \eref{eq:W_ij_V} & $-2.14\ten{-5}$ &
  \eref{eq:W_qbarq_V_(4)} \\ \hline
$g+g\rightarrow Z+b+\bar{b}$ & $4.71\ten{-3}$ & \eref{eq:W_ij_V} &
 $4.56\ten{-3}$ & \eref{eq:W_gg_V_(1)}
 \\ \hline \hline
\multicolumn{5}{|c|}{$\sqrt{s} = 16\,{\rm TeV}$} \\
 & $\sigma_{\rm exact}^Z$ (nb) & & $\sigma_{\rm app.}^Z$ (nb) & \\ \hline
$q+\bar{q}\rightarrow Z+b+\bar{b}$ & $9.54\ten{-2}$ & \eref{eq:W_qbarq_V_2},
  \eref{eq:W_ij_V} & $8.01\ten{-2}$ & \eref{eq:W_qbarq_V_(1)},
  \eref{eq:W_qbarq_V_(2)} \\ \cline{2-5}
 & $5.93\ten{-4}$ & \eref{eq:W_ij_V} & $5.97\ten{-4}$ &
  \eref{eq:W_qbarq_V_(3)} \\ \cline{2-5}
 & $-6.88\ten{-6}$ & \eref{eq:W_ij_V} & $-2.30\ten{-5}$ &
  \eref{eq:W_qbarq_V_(4)} \\ \hline
$g+g\rightarrow Z+b+\bar{b}$ & $5.85\ten{-1}$ & \eref{eq:W_ij_V} &
 $5.77\ten{-1}$ & \eref{eq:W_gg_V_(1)}
 \\ \hline
\end{mytable}
\clearpage
\newpage
\section*{Figure captions}
{\bf Fig. 1} Two-loop graph, containing the heavy flavors in the
             triangular subloop, which contributes to the subprocess
             $q+\bar{q}\rightarrow Z$ \eref{eq:W_qbarq_Z_2}.\\[2ex]
{\bf Fig. 2} Two-loop graphs, containing the heavy flavors in the
             gluon self energy, which contributes to the subprocess
             $q+\bar{q}\rightarrow V$ \eref{eq:W_qbarq_V_2}.\\[2ex]
{\bf Fig. 3} One-loop graph with the heavy flavors in the triangle
             contributing to the subprocess $q+\bar{q}\rightarrow Z+g$
             \eref{eq:W_qbarq_Z_1}.\\[2ex]
{\bf Fig. 4} One-loop graph with the heavy flavors in the triangle
             contributing to the subprocess $g+q(\bar{q})\rightarrow
             Z+q(\bar{q})$ \eref{eq:W_qg_Z}.\\[2ex]
{\bf Fig. 5} Diagrams contributing to the subprocess $q+\bar{q}
             \rightarrow V+Q_1+\bar{Q}_2$; a, b: initial state radiation;
             c, d: final state radiation \eref{eq:W_ij_V}.\\[2ex]
{\bf Fig. 6} Diagrams contributing to the subprocess $g+g\rightarrow
             V+Q_1+\bar{Q}_2$ \eref{eq:W_ij_V}.\\[2ex]
{\bf Fig. 7} The heavy flavor content, represented by $R$
             \eref{eq:ratio_R}, of the cross section $d\sigma/dQ^2$
             for the processes: $p+\bar{p}\rightarrow\gamma^\ast+``X"$
             $\sqrt{s} = 0.63\,{\rm TeV}$ (solid line), $\sqrt{s}
             = 1.8\,{\rm TeV}$ (dotted line) and $p+p\rightarrow
             \gamma^\ast+``X"$ $\sqrt{s}=16\,{\rm TeV}$ (dashed line).
\newpage
\begin{figure}
\diagram{
  .8 .8 Scale Init [0 -3] Goto Init
  8 Topology
  [4 3] + Fermion [3 2] + Fermion [2 1] + Fermion
  [2 5] + Gluon [3 6] Gluon
  [6 7] Fermion [7 5] Fermion [5 6] Fermion
  [7 8] VectorBoson
  1 Goto 1 180 RGoto (q) Overline
  8 Goto ( Z) PutText
  4 Goto 1 180 RGoto (q) PutText
  6 Goto 1 -5 RGoto (Q) PutText
}
\vspace*{4cm}
\caption{\label{fig:W_qbarq_Z_2}}
\end{figure}
\begin{figure}
\diagram{
  .8 .8 Scale Init [0 -3] Goto Init
  11 Topology
  [2 3] + Fermion [3 7] + Fermion [7 4] + Fermion [4 1] + Fermion
  [4 5] Gluon [3 6] + Gluon [7 8] VectorBoson
  [6 5] 180 + FermionArc [5 6] 180 + FermionArc
  8 Goto ( V) PutText
  1 Goto 1 180 RGoto (q) Overline (2) <202020> - PutText
  2 Goto 1 180 RGoto (1) (q ) - PutText
  8 Goto 5.5 180 RGoto (Q) PutText
  [12 0] Goto Init
  14 Topology
  [1 3] - Fermion [3 4] - Fermion [4 7] - Fermion
  [2 7] + Fermion [7 8] + VectorBoson
  [5 6] 180 + FermionArc [6 5] 180 + FermionArc
  [3 5] 90 GluonArc [6 4] 90 GluonArc
  8 Goto ( V) PutText
  5 Goto 1.2 60 RGoto (Q) PutText
  1 Goto 1 180 RGoto (q) Overline (2) <202020> - PutText
  2 Goto 1 180 RGoto (1) (q ) - PutText
}
\vspace*{9.5cm}
\caption{\label{fig:W_qbarq_V_2}}
\end{figure}
\begin{figure}
\diagram{
  .8 .8 Scale Init [0 -3] Goto Init
  [16 0] Goto
  10 Topology
  [2 3] + Fermion [3 1] + Fermion
  [3 4] Gluon
  [6 5] Fermion [5 4] Fermion [4 6] Fermion
  [5 7] + Gluon [6 8] VectorBoson
  8 Goto ( Z) PutText
  1 Goto 1 180 RGoto (q) Overline
  2 Goto 1 180 RGoto (q) PutText
  4 Goto 1 90 RGoto (Q) PutText
}
\vspace*{5cm}
\caption{\label{fig:W_qbarq_Z_1}}
\end{figure}
\begin{figure}
\diagram{
  .8 .8 Scale Init [0 -3] Goto Init
  [9 0] Goto Init
  9 Topology
  [1 2] + Fermion [2 3] + Fermion
  [2 4] + Gluon [7 5] + Gluon
  [4 6] Fermion [6 5] Fermion [5 4] Fermion
  [6 8] VectorBoson
  8 Goto ( Z) PutText
  1 Goto 1 180 RGoto (q) PutText
  3 Goto ( q) PutText
  4 Goto 1 145 RGoto (Q) PutText
}
\vspace*{5.5cm}
\caption{\label{fig:W_qg_Z}}
\end{figure}
\begin{figure}
\diagram{
  .8 .8 Scale Init [0 -2] Goto Init Save
  6 Topology
  [4 3] + Fermion [3 2] + Fermion [2 1] + Fermion
  [3 6] Gluon [8 6] + Fermion [6 7] + Fermion
  [2 5] + VectorBoson
  1 Goto 1 180 RGoto (q) Overline (2) <202020> - PutText
  4 Goto 1 180 RGoto (1) (q ) - PutText
  5 Goto ( V) PutText
  7 Goto ( Q) PutText
  8 Goto ( ) PutText (Q) Overline
  [3.5 -5] Goto (a) PutText
  Restore [11 0] Goto Init
  7 Topology
  [4 3] + Fermion [3 2] + Fermion [2 1] + Fermion
  [2 6] Gluon [8 6] + Fermion [6 7] + Fermion
  [3 5] VectorBoson
  1 Goto 1 180 RGoto (q) Overline (2) <202020> - PutText
  4 Goto 1 180 RGoto (1) (q ) - PutText
  5 Goto ( V) PutText
  7 Goto ( Q) PutText
  8 Goto ( ) PutText (Q) Overline
  [3.5 -5] Goto (b) PutText
  Restore [0 -8] Goto Init 1 Topology
  [2 3] + Fermion [3 1] + Fermion [3 4] Gluon
  [6 4] + Fermion [4 5] + Fermion [5 7] + Fermion
  [5 8] + VectorBoson
  1 Goto 1 180 RGoto (q) Overline
  2 Goto 1 180 RGoto (q) PutText
  8 Goto ( V) PutText
  7 Goto (1) ( Q ) - PutText
  6 Goto ( ) PutText (Q) Overline (2) <2020202020> - PutText
  [2.5 -5] Goto (c) PutText
  Restore [11 -8] Goto Init
  2 Topology
  [2 3] + Fermion [3 1] + Fermion [3 4] Gluon
  [7 6] + Fermion [6 4] + Fermion [4 5] + Fermion
  [6 8] VectorBoson
  1 Goto 1 180 RGoto (q) Overline
  2 Goto 1 180 RGoto (q) PutText
  8 Goto ( V) PutText
  5 Goto (1) ( Q ) - PutText
  7 Goto ( ) PutText (Q) Overline (2) <2020202020> - PutText
  [2.5 -5] Goto (d) PutText
}
\vspace*{11cm}
\caption{\label{fig:W_qbarq_V_QbarQ}}
\end{figure}
\begin{figure}
\diagram{
  .8 .8 Scale Init [0 -2] Goto Init Save
  1 Topology
  [1 3] + Gluon [2 3] + Gluon [3 4] Gluon
  [4 5] + Fermion [5 7] + Fermion [4 6] - Fermion
  [5 8] + VectorBoson
  8 Goto ( V) PutText
  7 Goto (1) ( Q ) - PutText
  6 Goto ( ) PutText (Q) Overline (2) <2020202020> - PutText
  Restore [11 0] Goto Init
  2 Topology
  [1 3] + Gluon [2 3] + Gluon [3 4] Gluon
  [4 5] + Fermion [4 6] - Fermion [6 7] - Fermion
  [6 8] VectorBoson
  8 Goto ( V) PutText
  5 Goto (1) ( Q ) - PutText
  7 Goto ( ) PutText (Q) Overline (2) <2020202020> - PutText
  Restore [0 -6] Goto Init
  3 Topology
  [1 2] Gluon [6 7] Gluon
  [2 6] - Fermion [2 3] + Fermion [3 4] + Fermion [6 8] - Fermion
  [3 5] + VectorBoson
  5 Goto ( V) PutText
  4 Goto (1) ( Q ) - PutText
  8 Goto ( ) PutText (Q) Overline (2) <2020202020> - PutText
  Restore [8 -6] Goto Init
  4 Topology
  [1 2] Gluon [6 7] Gluon
  [8 4] + Fermion [4 6] + Fermion [6 2] + Fermion [2 3] + Fermion
  [4 5] VectorBoson
  5 Goto ( V) PutText
  3 Goto (1) ( Q ) - PutText
  8 Goto ( ) PutText (Q) Overline (2) <2020202020> - PutText
  Restore [16 -6] Goto Init
  5 Topology
  [1 2] Gluon [6 7] Gluon
  [8 6] + Fermion [6 4] + Fermion [4 2] + Fermion [2 3] + Fermion
  [4 5] VectorBoson
  5 Goto ( V) PutText
  3 Goto (1) ( Q ) - PutText
  8 Goto ( ) PutText (Q) Overline (2) <2020202020> - PutText
  Restore [0 -16] Goto Init
  3 Topology
  [1 2] Gluon [6 7] Gluon
  [2 6] + Fermion [2 3] - Fermion [3 4] - Fermion [6 8] + Fermion
  [3 5] VectorBoson
  5 Goto ( V) PutText
  8 Goto (1) ( Q ) - PutText
  4 Goto ( ) PutText (Q) Overline (2) <2020202020> - PutText
  Restore [8 -16] Goto Init
  4 Topology
  [1 2] Gluon [6 7] Gluon
  [8 4] - Fermion [4 6] - Fermion [6 2] - Fermion [2 3] - Fermion
  [4 5] VectorBoson
  5 Goto ( V) PutText
  8 Goto (1) ( Q ) - PutText
  3 Goto ( ) PutText (Q) Overline (2) <2020202020> - PutText
  Restore [16 -16] Goto Init
  5 Topology
  [1 2] Gluon [6 7] Gluon
  [8 6] - Fermion [6 4] - Fermion [4 2] - Fermion [2 3] - Fermion
  [4 5] VectorBoson
  5 Goto ( V) PutText
  8 Goto (1) ( Q ) - PutText
  3 Goto ( ) PutText (Q) Overline (2) <2020202020> - PutText
  }
\vspace*{18.5cm}
\caption{\label{fig:W_gg_V_QbarQ}}
\end{figure}
\begin{figure}
  \epsfxsize=450pt
  \epsfysize=450pt
  \epsffile{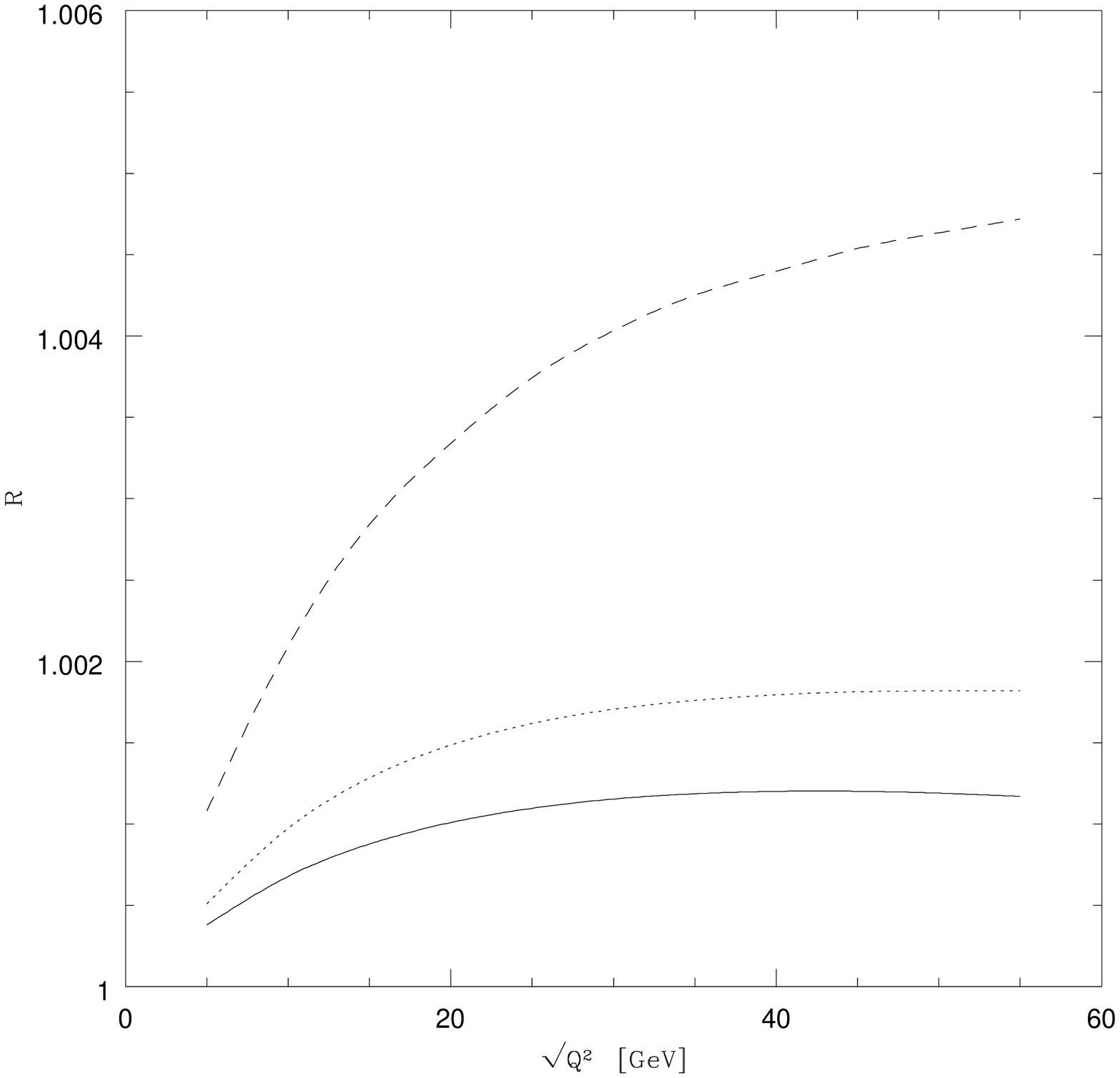}
  \caption{\label{fig:photo_prod}}
\end{figure}
\clearpage
\end{document}